\documentclass[esd, ms]{copernicus}

\begin{document}\sloppy

\title{No way out? The double-bind in seeking global prosperity alongside mitigated climate change}

\author{T.~J.~Garrett}

\affil{Department of Atmospheric Sciences, University of Utah, Salt Lake City, Utah, USA}

\runningtitle{Coupled evolution of the economy and the atmosphere}
\runningauthor{T.~J.~Garrett}

\correspondence{T.~J.~Garrett (tim.garrett@utah.edu)}

\received{21 March 2011}
\accepted{31 March 2011}

\published{}

\firstpage{1}

\maketitle

\begin{abstract}
In a prior study \citep{GarrettCO2_2009}, I introduced a simple
economic growth model designed to be consistent with general thermodynamic laws. Unlike traditional economic models, civilization is viewed only as a well-mixed global whole with no distinction made between individual nations, economic sectors, labor, or capital investments. At the model core is a hypothesis that the global economy's current rate of primary energy consumption is tied through a constant to a very general representation of its historically accumulated wealth. Observations support this hypothesis, and indicate that the constant's value is $\lambda$\,=\,9.7\,$\pm$\,0.3 milliwatts per 1990 US dollar. It is this link that allows for treatment of seemingly complex economic systems as simple physical systems. Here, this growth model is coupled to a linear formulation for the evolution of globally well-mixed atmospheric CO$_{2}$ concentrations. While very simple, the coupled model provides faithful multi-decadal hindcasts of trajectories in gross world product (GWP) and CO$_{2}$. Extending the model to the future, the model suggests that the well-known IPCC SRES scenarios substantially underestimate how much CO$_{2}$ levels will rise for a given
level of future economic prosperity. For one, global CO$_{2}$ emission rates cannot be decoupled from wealth through efficiency gains. For another, like a long-term natural disaster, future greenhouse warming can be expected to act as an inflationary drag on the real growth of global wealth. For atmospheric CO$_{2}$ concentrations to remain below a ``dangerous'' level of 450\,ppmv \citep{HansenDangerous2007}, model forecasts suggest that there will have to be some combination of an unrealistically rapid rate of energy decarbonization and nearly immediate reductions in global civilization wealth.
Effectively, it appears that civilization may be in a double-bind. If civilization does not
collapse quickly this century, then CO$_{2}$ levels will likely end up
exceeding 1000\,ppmv; but, if CO$_{2}$ levels rise by this much, then the
risk is that civilization will gradually tend towards collapse.
\end{abstract}

\introduction

Despite decades of public awareness of the potential for fossil fuel
consumption to lead to dangerous climate change, anthropogenic emissions of
CO$_{2}$ have accelerated \citep{Canadell2007,Raupach2007}. The implications
of civilization continuing on this path are environmental changes that are
both irreversible and profound, including amplified hydrological extremes,
storm intensification, sea level rise, and extreme mammalian heat stress
\citep{HansenDangerous2007,AllanSoden2008,Solomon2009,Vermeer2009,Sherwood2010}.

The economic costs associated with addressing and coping with climate warming are normally quantified by coupling a system of economic equations to a medium
complexity climate model. Normally, these Integrated Assessment Models (IAMs)
make regionally-based assessments of the economics of production, investment,
consumption, welfare, discount rates, population and rates of technological
change. These economic functions are coupled to functions for atmospheric
temperature and climate damage. From within a parameter space that might be of order 100
variables, the model outcome is a long-term optimized trajectory for
long-term societal welfare to which policy measures (for example the Copenhagen Accord) can be compared \citep{Nordhaus2000,Keller2004,Nordhaus2010}. Uncertainty in the
optimal path, when addressed, is modeled using Monte Carlo simulations within
a portion of the total parameter space \citep{Mastrandrea2004}.

Modern IAMs are normally based on mainstream neo-classical economic growth models that, unlike climate models, do not explicitly represent flows as a material flux down
pressure gradients. Economic flows are allowed to become progressively decoupled from energy consumption and CO$_2$ emissions through gains in energy efficiency. Several of the widely used IPCC SRES scenarios even go so far as to allow economic growth to continue while CO$_2$ emissions stabilize or decline \citep{Raupach2007}. 

This ``{}have our cake and eat it too'' viewpoint has been disputed by many ecological economists. The argument against decoupling is that consumption
of energy is thermodynamically required for any system to evolve,
and there is no physical reason that the human system should be treated as an exception
\citep{Lotka1922,Georgescu-Roegen,Ayres2003}. Some have even suggested that policies aimed at improving energy efficiency might backfire through what is known as ``{}Jevons' Paradox'': energy is useful, and for a given level of resource availability, efficiency gains make it cheaper and more desirable, ultimately leading to greater rates of energy consumption and CO$_2$ emissions \citep{Saunders2000, Alcott2005,Owen2010, Alcott2010}.

This article continues in a similar conceptual vein, but it differs by treating the human system in a more strictly physical fashion. Here, no internal resolution is made of political divisions or economic sectors. Rather, civilization is treated only as a whole since internal economic trade and atmospheric mixing of CO$_2$ are very rapid compared to the multi-decadal evolution of civilization. Further, no explicit account is made of people or their policies. Civilization is part of the physical universe and it is modeled as any other physical system. Long-term growth in global consumption and emission rates are considered only as a thermodynamic response to civilization's expansion into newly available energy resources. 

Thus, unlike IAMs, this article does not evaluate what long-term policy actions will enable us to limit CO$_2$ emissions while maximizing global economic wealth. Rather, the aim is to explore the range of future trajectories that is actually physically possible: political will can only go as far as physical laws allow. The argument that will be presented is that, unfortunately, wealth cannot be decoupled from resource consumption. In fact, at least at the global scales that are relevant to CO$_2$ emissions, it appears that ``{}Jevons' Paradox'' does indeed apply: efficiency gains will backfire. For this reason, it is likely that all SRES scenarios considerably overestimate the extent of economic health that is possible for a given future atmospheric concentration of CO$_2$. Either global warming acts as an inflationary drag on the production of wealth; or, economic growth is sustained and atmospheric CO$_2$ concentrations accelerate their growth.

\section{A physically consistent economic framework\label{sec:Economic-framework}}

An earlier article introduced a simple macroeconomic
growth model that treats civilization in a manner consistent with physical conservation laws \citep{GarrettCO2_2009}. As illustrated in Fig.~\ref{fig:thermschematic},  all material within civilization is treated as being in local thermodynamic equilibrium with the same specific potential energy per unit matter; effectively, it is treated as a surface defined by constant temperature and pressure, constant density, or constant specific entropy. Accordingly, no distinction is made between the internal components of civilization. Unlike traditional economic models, no explicit account is made for labor, capital, households, firms, governments or banks, nor the flows to and
from these components. Rather, civilization is considered only as a whole, or at a sufficiently low resolution that the only resolved
distinction is between civilization and known primary energy reservoirs
(e.g.~coal, oil, uranium, etc.).

Flows to civilization can be viewed as a special case within the more general thermodynamic model shown in Figure ~\ref{fig:thermschematic}, a perspective that bears some similarities with the thermodynamic model used by \citet{AnnilaSalthe2009} to represent economic flows. Energy reservoirs lie along a higher potential surface than the system of interest. The interface that separates these two surfaces is defined by a fixed step in specific potential energy $\Delta\mu$ (units potential energy per material unit) and a number of material units defining the length of the interface $\breve{n}$. The total potential difference that is available to drive downward flows is the product of these two quantities, i.e.,  ${\Delta}G$\,=\,$\breve{n}\Delta\mu$.  The flow redistributes the overall balance of potential energy towards the lower potential surface. Total material is conserved as required by the First Law of Thermodynamics, and the flow is downhill as required by the Second Law of Thermodynamics. The flow represents a ``heating'' of the lower potential system. The heating sustains this open system against a nearly equal dissipative flow due to the loss of energy to the system's surroundings. 

For civilization, the heating is equivalent to the rate~$a$ (units energy per time) at which civilization consumes the potential energy in primary energy resources.  The flow rate of energy is proportional to the material length of the interface $\breve{n}$ through 
\begin{equation}
a~=~\alpha~{\Delta}G~=~\alpha~\breve{n}\Delta\mu\label{eq:aalphaG}
\end{equation}
where, $\alpha$ is a constant rate coefficient with units inverse time (effectively a diffusivity). This consumption of potential energy is more precisely defined as a material flux. For civilization, coal and oil are examples of the agents that carry the potential energy we consume. However, civilization is not made of coal and oil, but rather of raw materials such as iron and copper. Potential energy consumption enables these raw materials to be carried downward along with the energetic flow to add to civilization's material bulk and sustain it against decay.

If civilization's economic activities are part of the physical universe, then perhaps there might be a fiscal representation for the physical flows that sustain it. A hypothesis can be proposed that the size of civilization is expressible thermodynamically by the potential difference $\Delta G$ driving flows, or equivalently the heating of civilization at rate $a=\alpha\Delta G$. Since heating sustains all of civilization's activities against its ultimate dissipative loss to its surroundings, the heating rate might conceivably be what civilization intrinsically values, and therefore it might be related to a very general expression of civilization's real, or inflation-adjusted economic wealth through\begin{equation}
a~\equiv~{\lambda}C\label{eq:alamC}
\end{equation}
where the rate of
consumption of the potential energy in primary energy resources $a$~ (units energy
per time) is related through a constant parameter $\lambda$ to a
fiscal representation of global economic wealth~$C$ (units inflation-adjusted
currency). If there is no energy consumption, then civilization is worthless because the activities that sustain civilization's value cannot be maintained against civilization's energy loss through decay. Effectively civilization becomes indistinguishable from its surroundings because the interface $\breve{n}$ and the gradient $\Delta{G}$ shrink to zero. We eat to sustain ourselves against radiative heat loss. If we don't eat, eventually we die.

Here, wealth~$C$ is defined as the historical accumulation of
gross world economic real production~$P$ (units inflation-adjusted currency
per time). A comparison of this definition with more traditional approaches is presented in Section 4. Here, real \mbox{production~$P$} is an instantaneous quantity that is related
to the familiar gross world product (GWP) through
\begin{equation}
{\rm GWP}~=~{P}\Delta{t}\label{eq:GWP}
\end{equation}
where, for the sake of economic
statistics, ${\Delta}t$ is normally equal to one year. Total economic wealth
is distinct from production in that it is not a differential but an integral
quantity (units inflation-adjusted currency). As wealth is defined here, it
is represented by the historical accumulation of production
\begin{equation}
C~\equiv~\int_{0}^{t}~P~\left(t'\right)~dt'~\simeq~\sum_{i}~{\rm GWP}~\left(i\right)\label{eq:C}
\end{equation}
where $i$~is an index covering the full historical record for GWP. Equivalently, economic production is a consequence of a convergence of the material and energetic flows associated with wealth  
\begin{equation}
\frac{dC}{dt}~\equiv~P\label{eq:dCdtP}
\end{equation}
or, expressed thermodynamically, from Eqs. \ref{eq:aalphaG} and \ref{eq:alamC}
\begin{equation}
P~=~\frac{1}{\lambda}~\frac{da}{dt}~=~\Delta\mu\frac{\alpha}{\lambda}\frac{d\breve{n}}{dt}\label{eq:Palphaw}
\end{equation}
Effectively, economic production~$P$ is a fiscal representation of the growth rate
of energy consumption \textit{da/dt} through an expansion of civilization's material interface $\breve{n}$ into the primary energy reservoirs that it consumes. Combining Eqs.~(\ref{eq:C}) and~(\ref{eq:Palphaw}),  global wealth arises from an accumulation of a net material convergence over time:
\begin{equation}
C~\equiv~\frac{1}{\lambda}~\int_{0}^{t}~\frac{da}{dt}~dt'~=~\Delta\mu\frac{\alpha}{\lambda}~\int_{0}^{t}~\frac{d\breve{n}}{dt}~dt'\label{eq:Cw}
\end{equation}

Eqs.~(\ref{eq:aalphaG}) and~(\ref{eq:alamC}) imply a direct proportionality between wealth~$C$, rates of primary energy consumption~$a$, and the size of the interface driving flows  $\Delta{G}=\breve{n}\Delta\mu$. In this case, there is a rate of return $\eta$ that applies equally to each:
\begin{equation}
\eta~\equiv~\frac{d~\ln~{\Delta}{G}}{dt}~=~\frac{d~\ln~\breve{n}}{dt}~=~\frac{d~\ln~a}{dt}~=~\frac{d~\ln~C}{dt}\label{eq:rateofreturn}
\end{equation}
Positive values of $\eta$ allow for exponential growth associated with interface expansion. Civilization wealth and energy consumption are in exponential decay if the interface $\breve{n}$ shrinks. 

Thus, from Eqs.~(\ref{eq:dCdtP}) and~(\ref{eq:rateofreturn}), the economic production function for this framework is 
\begin{equation}
P~\equiv~\frac{dC}{dt}~=~{\eta}C\label{eq:PetaC}
\end{equation}
The rate of return $\eta$ (units inverse time) is a time varying quantity that relates the past accumulation of wealth $C$ to the current production of new wealth $P$. Finally, by taking the time derivative of Eq.~(\ref{eq:PetaC}), the GWP growth rate is given by
\begin{equation}
\frac{d~\ln~P}{dt}~=~\eta~+~\frac{d~\ln~\eta}{dt}\label{eq:dlogPdt}
\end{equation}
Thus, what is normally termed as ``economic growth'' (i.e.~$d{\ln}P/dt$) is related to the sum of the growth rate of energy consumption $\eta$ and the \textit{acceleration} of growth in energy consumption $d\ln\eta /dt$. The economic growth rate stalls if this acceleration stagnates.

\section{Model validation}

The above discussion rests on an assumed constancy of the parameter $\lambda$, as it is defined through Eqs.~(\ref{eq:alamC})
and~(\ref{eq:C}) by
\begin{equation}
\lambda~\equiv~\frac{a\left(t\right)}{\int_{0}^{t}~P\left(t'\right)~dt'}~\simeq~\frac{a~\left(t\right)}{\sum_{i}~{\rm GWP}~\left(i\right)}
\label{eq:lambda}\end{equation}

To evaluate the validity of a hypothetical constancy of $\lambda$ in Eq.~(\ref{eq:lambda}), I employed statistics for world
GWP spanning more than 2000~years \citep{Maddison2003,UNstats} to calculate
wealth~$C$ from Eq.~(\ref{eq:C}). Values of~$C$ were compared to nearly four
decades of statistics for energy consumption rates~$a$ \citep{AER2009}.

Details are described in Appendix C of \citet{GarrettCO2_2009}. As illustrated in Table~\ref{tab:Measured-values}, the comparison supports
the hypothesis that the value of $\lambda$, as defined by Eq.~(\ref{eq:lambda}),
is indeed a constant that is independent of time: energy
consumption rates~$a$ and wealth $C$\,=\,$\int_{0}^{t}Pdt'$ both approximately
doubled in tandem between 1970 and 2008. On a millennial scale this time
interval is short, but it covers a tripling of GWP and more than half of
total civilization growth. The full yearly time series indicates that, during
this period, $\lambda$ maintained a mean value, with associated 95\% confidence uncertainty
in the mean, of 9.7\,$\pm$\,0.3 milliwatts per 1990\,US dollar
\citep{GarrettCO2_2009}. 

A theoretically equivalent approach to calculating $\lambda$ is to take the respective derivatives of $a$ and $C$ in order to compare the inter-annual change in energy consumption rates $da/dt$ to the real GWP $P$ (Eq. \ref{eq:Palphaw}). Derivatives of measured quantities are always more noisy than their integrals. For example, the magnitude of $d\ln{a}/dt$ is only about a couple of percent per year, where $a$ itself is subject to measurement uncertainties that, while unpublished, are plausibly of a similar magnitude. Nonetheless, the calculated mean value of $P/(da/dt)$ for the 1970 to 2008 period is 11.6\,$\pm$\,4.1 milliwatts per 1990\,US dollar, which is statistically consistent with the derived value for $\lambda\equiv a/C$ of 9.7\,$\pm$\,0.3 milliwatts per 1990\,US dollar.

This combination of theoretical and observational support for there being a fixed relationship between $C$ and $a$ is the key result supporting this study.
It serves as the basis for assuming that civilization is financially
well-mixed and that wealth is derived most fundamentally from a capacity to
enable a flow of potential energy from primary energy reserves. If it is generally correct, it enables an enormous simplification of what is required to accurately model the global
economy and its waste products. At least at a global scale, a sophisticated
IAM approach that explicitly considers people and their lifestyles is not
necessary in order to forecast future rates of energy consumption. People do
not need to be explicitly resolved in order to calculate global scale
consumptive flows.

As a note, the constancy of $\lambda$ should not be expected to hold at national scales. One country could easily be relatively wealthy compared to its current rate of energy consumption, provided that other countries are relatively poor. The value of $\lambda$ is constant only as a global quantity, where $C$ and $a$ subsumes all countries that are connected through international trade.

\section{Comparison with traditional economic growth models}

The model presented here is unlike traditional models in several regards, but it also has key similarities (see also Appendix~B in \citet{GarrettCO2_2009}). Wealth~$C$ is analogous to the term ``capital'' used in traditional economic growth
frameworks in the sense that it has units of currency, rather than currency
per unit time. However, it is much more general.  As shown in
Figure~\ref{fig:thermschematic}, civilization is defined as a whole, and no
distinction is made between the human and non-human elements of the global
economic system. Economic elements are not independent. Rather, all economic elements in civilization form a generalized capital that works in concert to consume primary energy reserves and enable the ``downhill'' flows of material in a field of potential energy. 

Effectively, treating civilization as a whole means that it is assumed to be
internally at local thermodynamic equilibrium, homogeneous, or ``well-mixed''. This does not mean that all economic elements are equal in value (they are not), only that the speed of financial interactions between all civilization elements is sufficiently rapid compared to the
timescales of global economic growth that internal gradients can be ignored.

A consequence of treating civilization as a whole is that human labor is part of a more general expression of capital $C$. Traditional economic models
separate ``{}physical'' capital from labor as distinct motive forces of economic production
\citep{Solow1956}, sometimes including supplies of energy and raw materials in an appeal to thermodynamic constraints \citep{Saunders2000,WarrAyres2006}. Labor, capital and energy inputs are all set to exponents that are tuned to provide agreement with observed sectoral or national production statistics.  Capital grows only due to ``investments'' that are separated from household and government ``consumption''. Household consumption never adds to capital. For one, people are not normally considered to be part of capital. For another, value that is consumed is presumed to be gone forever, so consumption must be subtracted from production to obtain the true capital investment.

Here, however, humans are subsumed into capital so that the production function, given by $P = \eta C$ (Eq. \ref{eq:PetaC}), is determined only by the general expression of capital used here and a variable rate of return $\eta$ that might be analogous to the ``{}total factor productivity'' employed by \citet{Solow1956}. Consequently, human consumption cannot be selectively subtracted from the production of new capital because humans are part of the whole. The component of economic production that is traditionally termed consumption is in fact an investment in sustaining and growing humanity. 

That said, physically it makes most sense to refer to consumption as something that is much more extensive than what is directly tallied in economic accounts. In Figure~\ref{fig:thermschematic}, consumption is proportional to the global scale flow of primary energy resources as it passes \emph{through} civilization. This consumptive flow of matter and potential energy is downhill from high to low potential at right angles to the constant potential surface along which civilization lies. Economic production is proportional to the expansion of this potential surface. Thus, consumption and production cannot be differenced because the two quantities are mathematically orthogonal. Consumption is not a component of production, but rather production is the \emph{convergence} in thermodynamic consumption. Only if civilization as a whole consumes more energy than it dissipates can the interface expand and net economic value be produced.

An added advantage of subsuming labor into capital, where capital is fundamentally assumed to be an implicit representation of energy consumption through $a\equiv\lambda C$, is that, unlike traditional models, there is no need for any tuning of non-integer exponents in a production function. Tuning to prior data can be a useful tool of last resort. But, it has its problems because there is little guarantee that a model tuned to the past will not need retuning to be consistent with data in the future. While the physical approach discussed here may be highly unorthodox by mainstream economic standards, it does have the advantage that its absence of a tuning requirement allows it to rest on a testable, falsifiable hypothesis -- falsifiability is one of the key hallmarks of science. Either there exists a constant  coefficient $\lambda$, or there doesn't. Of course, as discussed above, the constancy in $\lambda$ does appear to hold. But the point is that if this constancy ever fails, then the model presented here can be safely dismissed. Retuning is not an option.

\section{Jevons' Paradox and why efficiency gains accelerate global CO$_2$ emission rates}

Certainly, it might seem obvious that technological advances that increase energy efficiency or energy productivity (defined as $P/a$) should lead to a decrease in CO$_{2}$ emissions. Less energy is required to accomplish the same economic task. Even so, there is recognition among many economists of the existence of a {}``rebound
effect'', whereby increasing energy productivity spurs greater emissions
further down the road \citep{Dimitropoulos2007,Herring2007,Sorrell_UKERC2007}.
Two types of rebound have been identified, although in essence they
both address the issue of what happens to whatever money is saved
when energy productivity goes up. The {}``direct'' rebound effect
is limited to a particular energy service. For example, people may
choose to drive more often if a vehicle is fuel efficient, because
driving is useful or pleasurable and now more affordable. There are
also {}``indirect rebound effects'', which extend response to differing
economic sectors. Less money spent on fuel for efficient vehicles
might enable more money to be spent on fuel for home heating.

A few studies even point to an extreme form of rebound termed {}``backfire'': gains in energy efficiency lead ultimately to more rather than less energy consumption \citep{Saunders2000, Alcott2005,Owen2010, Alcott2010}. First discussion of the
principle came from an 1865 exposition on energy economics by William
Stanley Jevons \citep{Jevons1865}. Jevons was emphatic that the
introduction of energy efficient steam engines had accelerated Britain's consumption of coal. The cost of steam-powered coal extraction became cheaper and, because coal was very useful, more attractive. 

While the topic has received revived attention politically \citep{HoL2006},
a general consensus on the total magnitude of the effect has proved
elusive \citep{Sorrell_UKERC2007}. One reason is that calculating
the knock-on effects from an efficiency improvement in one sector
as they propagate through the entire global economy is daunting if
not impossible. Suppose that efficient
vehicles enable added household heating through a savings in transportation costs. Then, by raising home comfort, workers sleep better so that they are better able to facilitate
resource extraction at their companies. With higher profits, the companies
then reward the workers with raises, who in turn spend the money on
goods produced overseas with coal-generated electricity. So, in this
fashion, the ramifications of any given efficiency action can multiply
indefinitely, spreading at a variety of rates throughout the global
economy. Barring global analysis of rebound effects over long time
scales, conclusions may be quantitative but uncertain, and dependent
on the time and spatial scales considered.

An easy way to address this problem is to not resolve economic flows within the global economy, but rather to take the more general approach shown in Fig. \ref{fig:thermschematic}. In this case, energy efficiency is defined only with respect to the economic capacity of civilization, taken as a whole, to grow by doing work on its surroundings, allowing it to expand into the reserves of primary energy that sustain it. The amount of net or real work that civilization does to grow itself depends on a balance between civilization's consumptive and dissipative flows. If civilization is efficient, there is a net material and energetic convergence that allows civilization to do net positive work to ``stretch'' outward its interface with respect to its primary energy supplies. If energy efficiency increases, this accelerates civilization expansion, allowing civilization to consume energy at an ever faster rate.

Expressed in analytical terms, consumption of primary energy resources at rate~$a$ enables work to be done at rate~$w$ in order to extend the material size $\breve{n}$ of the interface that drives consumptive flows. From Eq. \ref{eq:aalphaG},  work is done at rate
\begin{equation}
w~=~{\Delta}\mu\frac{d\breve{n}}{dt}~=~{\epsilon}a\label{eq:ddeltaGdt}
\end{equation}
where $\epsilon$\,=\,$w/a$ is the efficiency for the conversion of heat transfer to
work. Unlike the normal conception, where work is done only to raise the
potential of some outside agency, here work is more self-referential. Work is done by civilization to increase the size and consumptive capacity of civilization itself.

If net work is positive, then there is exponential growth in the rate of primary energy
consumption $a$. Interface expansion into new energy reservoirs creates a positive feedback loop by bootstrapping civilization into an ever more consumptive state. Combining Eqs.~(\ref{eq:aalphaG}) and~(\ref{eq:ddeltaGdt}), the rate of increase in energy consumption is related to the work done to expand the interface through
\begin{equation}
\frac{da}{dt}~=~\alpha~{\Delta}\mu\frac{d\breve{n}}{dt}~=~{\alpha}w\label{eq:dadt}
\end{equation}
where, as before, $\alpha$ is an unknown constant.  Since $w={\epsilon}a$, dividing by $a$ provides an expression for the ``rate of return'' on consumption $\eta$, as defined previously in Eq.  \ref{eq:rateofreturn}, that is directly proportional to energy efficiency through
\begin{equation}
\eta = \frac{1}{a}\frac{da}{dt}~=~\alpha\frac{ w}{a} = {\alpha}\epsilon\label{eq:etaefficiency}
\end{equation}
Thus, global scale increases in the energy efficiency $\epsilon$
lead to a higher rate of return $\eta$ and accelerated growth of energy consumption rates~$a$. Treated as a whole, an efficient system grows faster and consumes more. 

That said, increasing energy efficiency does translate to higher prosperity. Economic production is related to the rate of return through $P=\eta C$ (Eq. \ref{eq:PetaC}), where wealth $C$ is tied to energy consumption through $a = \lambda C$ (Eq. \ref{eq:alamC}), $\lambda$ being an empirically measurable constant. It follows that, at global scales, the energy productivity $P/a$ is tied to energy efficiency $\epsilon$ through 
\begin{equation}
\frac{P}{a}~=~\frac{\eta}{\lambda}~=~\frac{\alpha}{\lambda}\epsilon\label{eq:productionefficiency}
\end{equation}
The implication is that, at least for global economic systems, changes in energy efficiency and energy productivity are equivalent. Through Eq. \ref{eq:dlogPdt}, both accelerate GWP growth even if they do not in fact lead to a decrease in overall energy consumption, as is commonly assumed \citep{PacalaSocolow2004,Raupach2007}. At global scales, Jevons' Paradox holds.
 
The analogy here might be to a growing child, who uses the material nutrients and
potential energy in food not only to produce waste but also to grow. As the child
grows, it eats more food, accelerating its growth until it reaches adulthood
and growth stabilizes (in which case $\eta$\,$\simeq$\,0). A healthy, energy
efficient child will grow faster than one who is sick and inefficient. A
diseased child may even die (in which case $\eta$\,$<$\,0).

These conclusions have direct bearing on global scale emissions of CO$_{2}$. Just as civilization can be treated as
being well-mixed over timescales relevant to economic growth, atmospheric
concentrations of CO$_{2}$ are also well-mixed over timescales relevant to
global warming forecasts. Thus, for the purpose of relating the economy to
atmospheric CO$_{2}$ concentrations, what matters is only how fast
civilization as a whole is emitting CO$_{2}$. 

CO$_{2}$ emissions are primarily a by-product of energy combustion. The emission rate $E$ is determined by the product of the global rate of energy consumption~$a$, and the carbonization of the fuel supply defined by
\begin{equation}
c~\equiv~\frac{E}{a}\label{eq:Eca}
\end{equation}
where, $E$ and $a$ are
measured quantities. It follows from
Eq.~(\ref{eq:alamC}) that current rates of CO$_{2}$ emissions $E$ are fundamentally
coupled to wealth~$C$, or past economic production, through
\begin{equation}
E~=~{\lambda}cC\label{eq:ElamcC}~=~{\lambda}c{\int_{0}^{t}~P\left(t'\right)~dt'}
\end{equation}
Drawing from statistics for CO$_{2}$ emissions from the Carbon Dioxide Information
Analysis Center \citep{Marland2007}, Table~\ref{tab:Measured-values} shows
that, like $a$ and $C$, CO$_{2}$ emissions $E$ have approximately doubled
between 1970 and 2008. Meanwhile, the value ${\lambda}c$\,=\,$E/C$ has stayed
approximately constant. Its mean value (and uncertainty in the mean) taken
from the entire yearly time series is 2.42\,$\pm$\,0.02\,ppmv\,atmospheric equivalent CO$_{2}$ per year, per thousand trillion 1990\,US dollars of global wealth.

Note that, unlike $\lambda$, the carbonization~$c$ is not a fundamental
property of the economic system within this framework. At least in principle,
it could be more variable in the future than it has been in the recent past.
Combining Eqs.~(\ref{eq:rateofreturn}) and~(\ref{eq:ElamcC}), emission rates grow
at rate that is determined by the growth rate of wealth and the rate of
change of carbonization
\begin{equation}
\frac{d~\ln~E}{dt}~=~\frac{d~\ln~C}{dt}~+~\frac{d~\ln~c}{dt}~=~\eta~+~\frac{d~\ln~c}{dt}\label{eq:dlnEdt}
\end{equation}

The implication is that, if technological changes allow energy productivity or energy efficiency to increase, then the rate of return $\eta$ increases and CO$_2$ emissions accelerate. This is unless decarbonization is as rapid as the rate of growth of wealth $\eta$. If so, then emission rates~$E$ can be stabilized. If, however, the carbonization~$c$ 
stays approximately constant, then CO$_{2}$ emissions rates~$E$ will remain
fundamentally linked to global economic wealth~$C$ through the constant value of 2.42\,$\pm$\,0.02\,ppmv\, of CO$_{2}$ emitted per year, per thousand trillion 1990\,US dollars of current wealth. It can only be through an economic collapse that CO$_{2}$ emissions rates will decline.

\section{Environmentally driven economic decay}

\subsection{Accounting of inflation and decay\label{economicdecay}}

The broadest available measure of the inflation rate is the so-called GDP deflator, which is calculated from the year-on-year increase in the prices of a very broad basket of consumer and industrial goods. Effectively, the gross domestic product becomes devalued
by some inflationary fraction~$i$ that makes the ``real'',
inflation-adjusted GDP less than its ``nominal'' value. Expressed for the world as a whole
\begin{equation}
i~=~\frac{{\rm nominal}~-~{\rm real}}{\rm nominal}~=~\frac{\hat{\rm GWP}~-~{\rm GWP}}{\hat{\rm GWP}}\label{eq:delta}
\end{equation}

While there have been a wide variety of theoretical explanations for
what drives inflation, the field is fluid and none have been solidly rejected  \citep{Parkin2008}. Price inflation is traditionally viewed as a simple imbalance between the monetary supply and economic output, and therefore mostly a matter for central bank control. What is less clear is why high inflation appears to have a negative effect on inflation-adjusted economic growth \citep{Sarel1996}. There are also external forces that can create the initial inflationary pressure, such as an external shock to primary energy supplies \citep{Bernanke1997}, and even climate change, which drives up food prices through adverse effects on crop production \citep{Lobell2011}. 

From the perspective of the model presented here, inflationary pressures can arise from either decreasing energy availability or increasing environmental disasters. This can be assessed because the real value or wealth of civilization is fixed to its current capacity to consume primary energy resources through the constant coefficient $\lambda$, which has a value of 9.7$\pm$0.3 milliwatts per inflation adjusted 1990 dollar: in 2008, 16.4 TW of power supported 1656 trillion 1990 US dollars of civilization worth. For interpreting inflation, this coefficient provides an anchor for assessing real economic worth, at least for civilization as a whole.

Supposing that natural disasters destroy the capacity of life and property to consume energy, civilization's real value declines while plausibly keeping the availability of currency largely intact. Alternatively, while banks do not actively destroy civilization's capacity to consume energy, they might be excessively loose with currency. If so, the real currency value attached to the existing capacity to consume energy becomes diluted by an excessive availability of new currency, while real wealth stays fixed. Whether banks or climate extremes initiate the action, in either case, inflation should be expected to follow as a consequence of an introduced imbalance between real and nominal value. The availability of currency becomes out of proportion with the true capacity of civilization to consume primary energy supplies.  

Real, inflation-adjusted wealth has been defined here by $C$\,=\,$\int_{0}^{t}Pdt'$
(Eq.~\ref{eq:alamC}) or equivalently, the instantaneous function \textit{dC/dt}\,$\equiv$\,$P$
(Eq.~\ref{eq:PetaC}), where $P$ is the inflation-adjusted production.
Here, in effect, all production is a differential addition to a
generalized measure of wealth, provided it is adjusted for inflation. This adjustment to the nominal (non-inflation-adjusted) production of
wealth $\hat{P}$ can be expressed as a sink of existing wealth
${\gamma}C$, where $\gamma$ represents the rate at which existing wealth is being destroyed or lost due to natural decay
\citep{GarrettCO2_2009}
\begin{equation}
\frac{dC}{dt}~\equiv~P~=~\hat{P}~-~{\gamma}C\label{eq:sinksource}
\end{equation}
Thus, the rate of decay is simply
\begin{equation}
\gamma~\equiv~\frac{\hat{P}~-~P}{C}~=~\frac{\hat{P}~-~P}{\int_{0}^{t}Pdt'}\label{eq:gamma}
\end{equation}
Similarly, the rate $\beta$ at which wealth~$C$ leads to nominal production
$\hat{P}$ can be defined by
\begin{equation}
\beta~\equiv~\frac{\hat{P}}{C}~=~\frac{\hat{P}}{\int_{0}^{t}Pdt'}\label{eq:beta}
\end{equation}
In this case, from Eq.~(\ref{eq:sinksource}), the growth of wealth can be
expressed as a balance between a source and a sink of wealth
\begin{equation}
\frac{dC}{dt}~=~\left(\beta~-~\gamma\right)~C\label{eq:dCdt}
\end{equation}
This is just an alternative expression for Eq.~(\ref{eq:PetaC}) with the rate of return
on wealth $\eta$ replaced by the difference between the coefficient of
nominal production $\beta$ and the coefficient of decay $\gamma$
\begin{equation}
\eta~=~\beta~-~\gamma\label{eq:eta}
\end{equation}

The advantage of applying this treatment is that it leads to a very simple expression for
an inflationary pressure~$i$ in Eq.~(\ref{eq:delta})
\begin{equation}
i~=~\frac{\int_{t}^{t+{\Delta}t}~\left(\hat{P}~-~P\right)~dt'}{\int_{t}^{t+\Delta
t}\hat{P}~dt'}~=~\frac{\int_{t}^{t+{\Delta}t}~\gamma~Cdt'}{\int_{t}^{t+\Delta{t}}~\beta~Cdt'}
~=~\frac{\left\langle\gamma\right\rangle}{\left\langle\beta\right\rangle}\label{eq:inflation}
\end{equation}
where brackets imply a mean over the time interval of calculation
${\Delta}t$, which is normally one year. Inflation is determined by
the balance between the coefficients $\beta$ and $\gamma$ of production and decay.\footnote{In practice, statistics for nominal and real GWP are normally
provided in current and fixed-year currency, respectively, and therefore are
in different units. Thus, for a given time period ${\Delta}t$ (say one year),
$\gamma$ can be calculated from differences in the logarithmic rate of
expansion for $\hat{P}$ and $P$, noting that $\ln\left(1+x\right)$\,$\simeq$\,$x$
\[
\gamma~=~\frac{\hat{P}~-~P}{C}~\simeq~\frac{P}{C}~\left[\frac{1}{P}~\frac{d\left(\hat{P}~-~P\right)}{dt}\right]~{\Delta}t~\simeq~\frac{P}{C}~\frac{d~\ln~\left(\hat{P}/P\right)}{dt}~{\Delta}t
\]
Effectively $\left[d~\ln~\left(\hat{P}/P\right)/dt\right]{\Delta}t$ is the
fractional inflation~$i$ over period ${\Delta}t$. Then, since
$\eta$\,=\,$P/C$, it follows that $\gamma$\,=\,$i\eta$ and $\beta$\,=\,$\eta$\,+\,$\gamma$\,=\,$\left(1+i\right)\eta$.}
If ${\Delta}t$ is one year, then the quantity $i{\Delta}t$ represents the difference between nominal
and real GWP.

If the coefficient of decay becomes greater than the coefficient of
production, such that $\gamma$\,$>$\,$\beta$, then from  Eq.~(\ref{eq:inflation}),
\textit{nominal} production $\hat{P}$ may be positive, but \textit{real}
production~$P$ is negative. Discussing negative real production would seem
unusual (or impossible) from a traditional economic perspective that is geared towards modeling growth. From the
more physical framework described here, it is simply a consequence of
environmentally driven decay being so large that there are economic hyper-inflationary pressures associated with a rate $i$\,=\,$\gamma/\beta$ that is greater than 100\%.  Historically, and on more regional levels, this is in fact
a fairly common scenario. From Eq.~(\ref{eq:sinksource}), \textit{dC/dt}\,$<$\,0, and total wealth is in a state of collapse. 

As discussed in Appendix A, hyper-inflation and collapse can be viewed thermodynamically as an interface between civilization and its energy reserves that is shrinking inwards rather than growing outwards. This means that the nominal work ${\int_{t}^{t+\Delta
t}\hat{w}~dt'}$ that is done to grow civilization is overwhelmed by external work done on civilization through decay. Real or net work done to grow civilization $\int_{t}^{t+\Delta{t}}{w}~dt'$ turns negative and civilization enjoys no return on its energetic investment. As a whole, civilization becomes less wealthy as it becomes less able to consume primary energy reserves.

A related concept is termed Energy Returned on Energy Invested (or EROI), and is becoming increasingly popular as a metric of society's capacity to exploit primary energy reserves for economic benefit \citep{Murphy2010}. Evidence suggests that the value of EROI is declining, presumably as new energy reserves become increasingly difficult to access. In Appendix A it is shown that a direct link can be drawn between the EROI concept and inflation (or the GDP deflator) discussed here. At global scales, the value of EROI is equal to the inverse of the inflation rate.

\subsection{Inflationary pressures and civilization resilience}

The IPCC Working Group~II
\citep{IPCC_WG22007} has identified potential societal damages due to climate ``extremes'', such as droughts and floods, and ``means'', such as
sea-level rise. These will exert a negative feedback on civilization wealth
such that, at some point, wealth and atmospheric CO$_{2}$ become
intrinsically coupled because civilization is no longer able to consume and emit as it has in the past.

Based on the above arguments, it is easy to see how natural disasters are
often expected to be inflationary since they represent an increase in the work done by the environment \emph{on} civilization. If the decay coefficient $\gamma$ suddenly
rises, then from Eq.~(\ref{eq:sinksource}), this expands the difference between
nominal and real production. From Eq.~(\ref{eq:inflation}), the shock
leads to inflation and less capacity to consume energy and emit CO$_{2}$.

An important point here is that, for inflationary pressures to take
hold, there must be an increase not just in total damages ${\gamma}C$, but in
the \textit{coefficient} of decay $\gamma$. Hurricane
damages along the Atlantic seaboard have risen over the past century, but not because of a long-term increase in hurricane intensity or frequency (i.e., $\gamma$). Rather, economic wealth $C$ has become increasingly concentrated at the coasts \citep{Pielke_hurr2008}. 

What seems reasonable is to expect that the decay rate $\gamma$ will in fact
change over coming decades due to the increasingly harmful effects of
global warming. Impacts will be regionally specific, but extremely difficult
to predict. In light of this, the approach taken here is to simplify
representation of the global economic impacts of climate change by defining a
global economic ``resilience'' to a logarithmic increase in atmospheric
CO$_{2}$ concentrations
\begin{equation}
\rho~=~1/\left(d\gamma/d~\ln~\left[\mathcal{\rm CO}_{2}\right]\right)\label{eq:resilience}
\end{equation}
If civilization's resilience is high, then the coefficient of decay $\gamma$
responds weakly to logarithmically changing CO$_{2}$ levels.\footnote{The
logarithm of CO$_{2}$ concentrations is considered because the primary
insulating gases responsible for climate warming, namely CO$_{2}$ and water
vapor, have a longwave absorptance that varies roughly as the square root of
their concentrations \citep{Liou-book}.}

\begin{figure}
\vspace*{2mm}
\center\includegraphics[width=12cm]{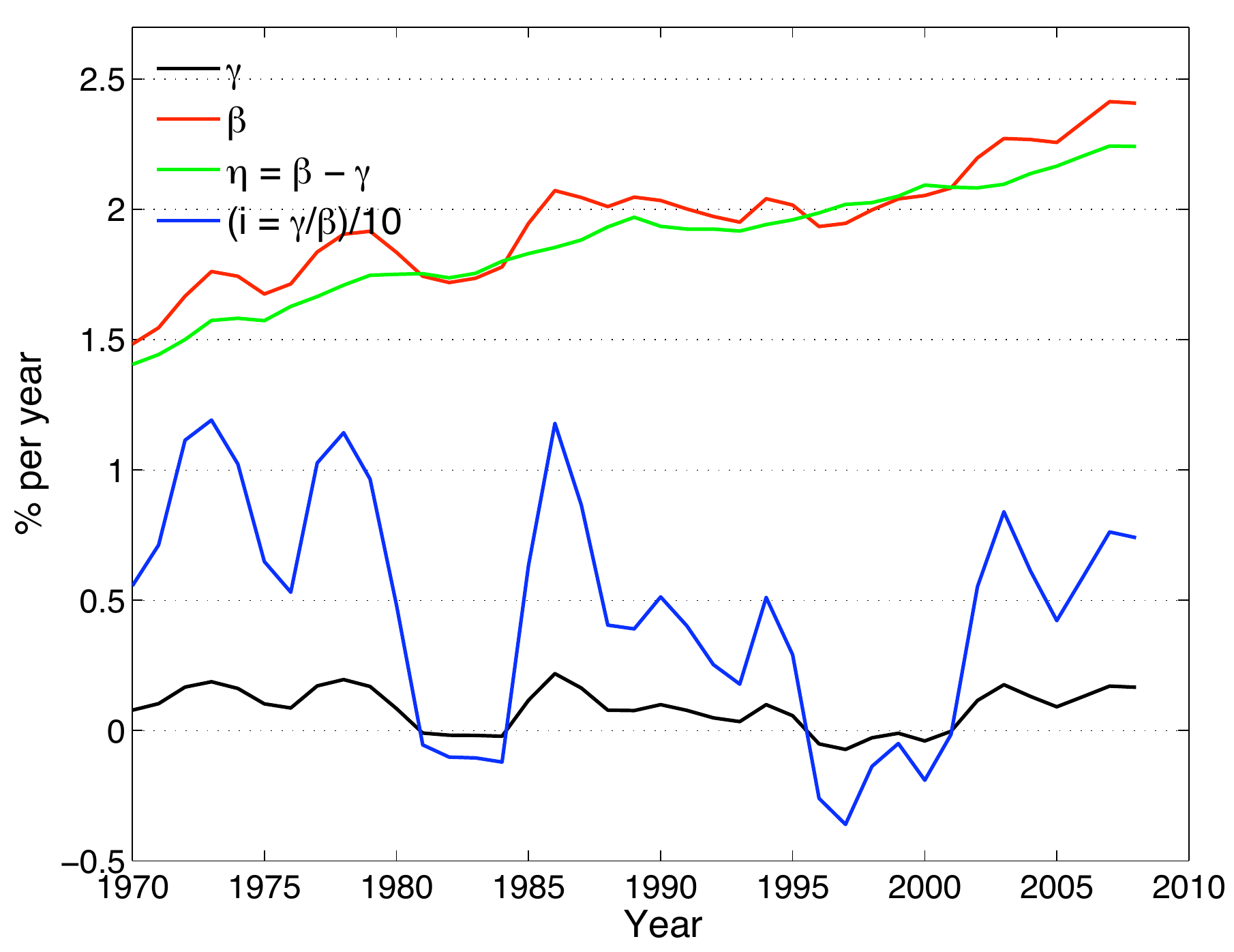}
\caption{From global economic statistics
\citep{UNstats}, derived global values for global inflation $i$ (Eq.~\ref{eq:inflation}),
the decay coefficient $\gamma$ (Eq.~\ref{eq:gamma}), the
source coefficient $\beta$ (Eq.~\ref{eq:beta}) and the rate of return $\eta$
(Eq.~\ref{eq:eta}) based on observations of nominal and real production, and
total global wealth.\label{fig:etabetagamma}}
\end{figure}

\begin{figure}
\vspace*{2mm}
\center\includegraphics[width=12cm]{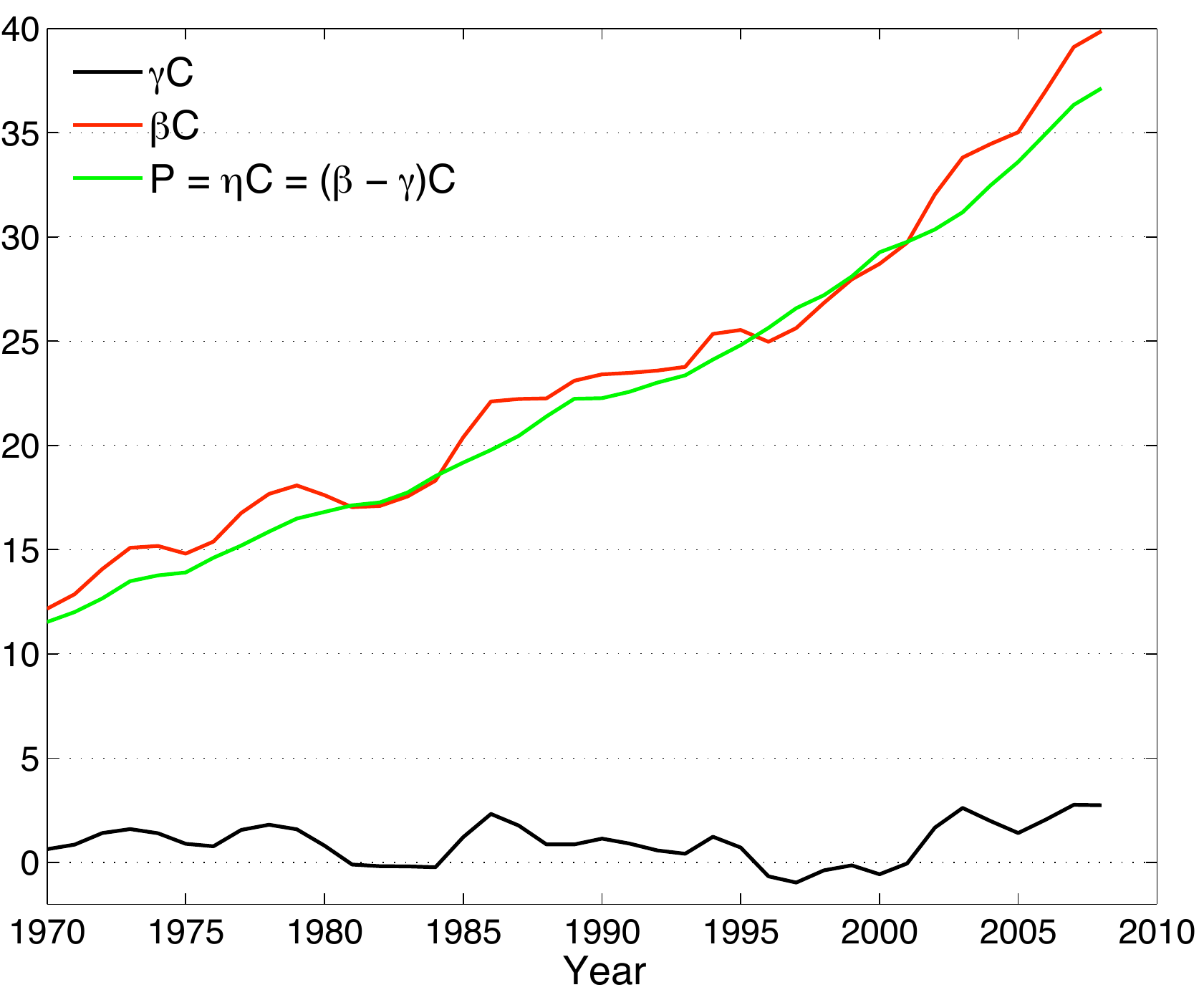}
\caption{As for Fig.~\ref{fig:etabetagamma}
but for the product of the rate coefficients and total wealth $C$ (Eq.~\ref{eq:C}).
The difference between ${\beta}C$ and ${\eta}C$ is the
inflationary depreciation associated with each year ${\gamma}C$. (Eqs.~\ref{eq:sinksource}
and~\ref{eq:eta}).\label{fig:etabetagammaC}}
\end{figure}

There have been estimates of the regional societal and economic impacts from extremes in
climate \citep{Patz2005,Leckebusch2007,Hsiang2011}. Unfortunately, it is not entirely obvious how to appropriately scale these impacts to civilization as a whole when many of the
effects of climate change will be sustained, global, and largely unprecedented. Recent
statistics do not yet provide meaningful guidance either. Figures~\ref{fig:etabetagamma}
and~\ref{fig:etabetagammaC} show no clear trends in the decay coefficient $\gamma$ that can easily be attributed to accelerating climate change. Up until this point, the dominant signature in $\gamma$ is only its inter-annual variability. A recent meta-analysis of disaster losses has arrived at a similar conclusion \citep{Bouwer2011}. 

The hypothesis that is proposed here is that the effect on society of
elevated levels of atmospheric CO$_{2}$ will be akin to a prolonged natural
disaster. From the standpoint of the economic model discussed above, the
effect will be to steadily increase the coefficient of decay $\gamma$ without
changing the coefficient of nominal production $\beta$. From Eq.~(\ref{eq:inflation}),
this will appear economically as an inflationary pressure
that impedes the growth in wealth $C$, as described by Eq.~(\ref{eq:dCdt}). In
a phase space of $\left[{\rm CO}_{2}\right]$ and $P$, the trajectory of
civilization will depend on the resilience $\rho$ of civilization to elevated
carbon dioxide levels: it is our resilience that will determine the strength
of climate's negative feedback on economic growth.

\section{The Climate and Thermodynamics Economic Response Model~(CThERM)}

\begin{figure}
\vspace*{2mm}
\center\includegraphics[width=12cm]{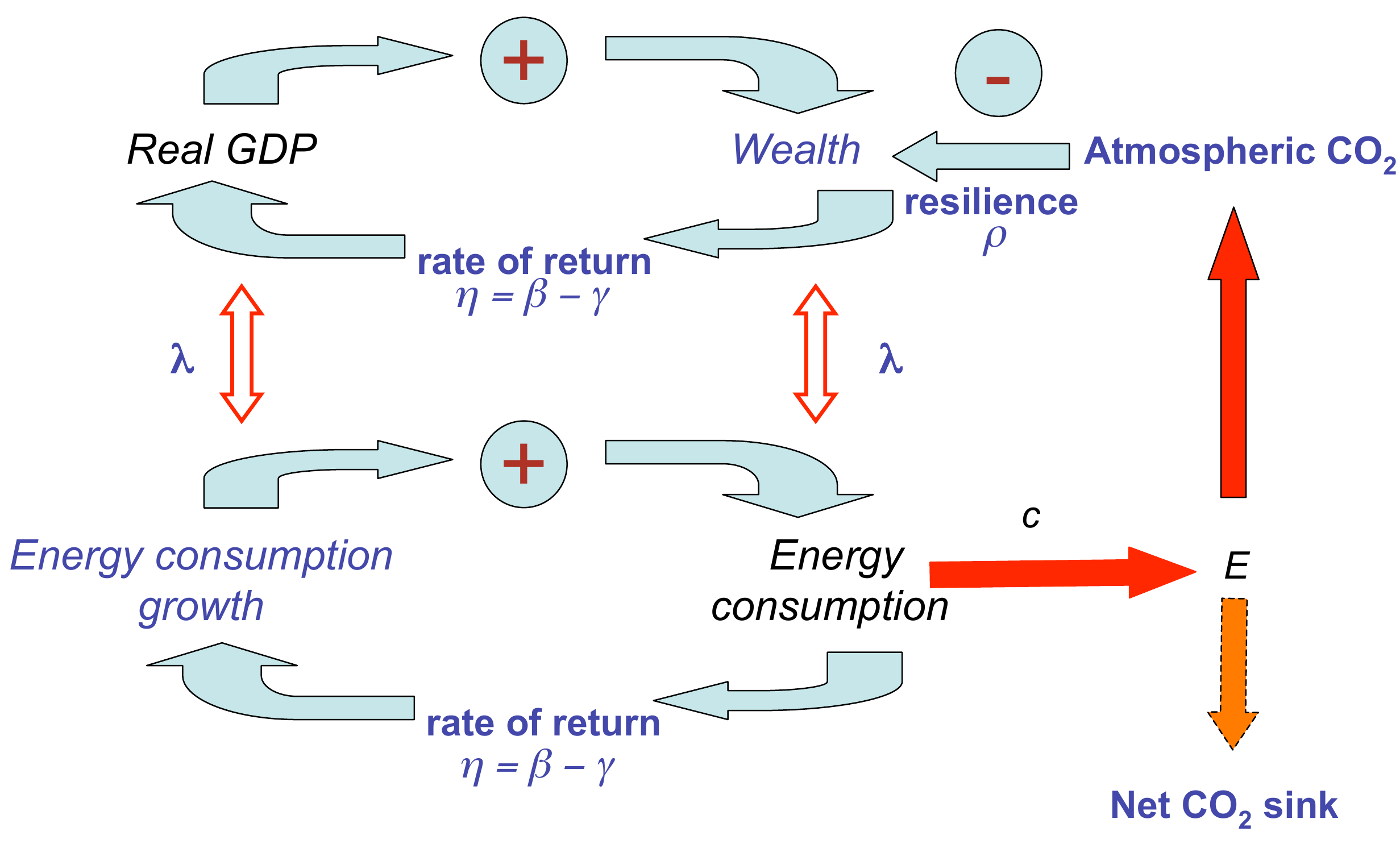}
\caption{Schematic illustrating the CThERM framework for
economic growth \citep{GarrettCO2_2009}, coupled to atmospheric CO$_{2}$
concentrations. Global rates of primary energy consumption rates~$a$ are tied
to accumulated inflation-adjusted global economic wealth $C$\,=\,$\int_{0}^{t}Pdt'$
through a constant coefficient $\lambda$\,=\,9.7 milliwatts per 1990~dollar.
Because $\lambda$ is a constant, growth in energy consumption rates
\textit{da/dt} are represented economically by the real, inflation-adjusted global GDP~$P$.
Thus, \textit{da/dt}\,=\,${\lambda}P$ determines the ``rate of return'' $\eta$\,=\,$d\ln\eta/dt$ adding to
$a$\,=\,${\lambda}C$. $E$ represents the anthropogenic rate of CO$_{2}$ emissions,
$\beta$ is the source for a positive rate of return $\eta$ due to increasing
availability of energy reservoirs. $\gamma$ is the sink for civilization
growth driven by environmental degradation. Emissions~$E$ determine CO$_{2}$
concentrations, subject to land and ocean sinks. CO$_{2}$ concentrations act
as a negative feedback on economic growth.\label{fig:EEaSM}}
\end{figure}

To explore the coupling between civilization and the atmosphere, the following section introduces a very simple framework for forecasting the
evolution of civilization in a phase space of $\left[{\rm CO}_{2}\right]$
and $P$, for a variety of assumed values of resilience $\rho$. The Climate and
Thermodynamics Economic Response Model (CThERM) couples a prognostic economic
model to atmospheric CO$_{2}$ concentrations, as illustrated in Fig.~\ref{fig:EEaSM}.
The prognostic economic module has just three coupled
dynamic equations for wealth~$C$, atmospheric CO$_{2}$ concentrations
{[}CO$_{2}${]}, and the rate of return $\eta$. From Eq.~(\ref{eq:rateofreturn}), wealth grows at rate
\begin{equation}
\frac{dC}{dt}~=~{\eta}C\label{eq:dlnCdt}
\end{equation}
 The balance between anthropogenic emissions $E$\,=\,${\lambda}cC$ (Eq.~\ref{eq:ElamcC})
 and natural sinks is
\begin{equation}
\frac{d~\left[{\rm CO}_{2}\right]}{dt}~=~E~-~\sigma~\Delta~\left[{\rm CO}_{2}\right]\label{eq:ddCO2dt}
\end{equation}
where $E$\,=\,${\lambda}cC$ (Eq.~\ref{eq:ElamcC}) and $\sigma$ is an assumed linear
sink rate on perturbations $\Delta\left[{\rm CO}_{2}\right]$\,=\,$\left[{\rm CO}_{2}\right]-\left[{\rm CO}_{2}\right]_{0}$ above some
preindustrial baseline.  For convenience, here it is assumed that the CO$_{2}$
emissions are instantly diluted in the total atmospheric mass
\citep{Trenberth1981} such that 1\,ppmv\,atmospheric CO$_{2}$\,=\,2.13\,Pg emitted carbon.
Thus $c$ is expressed in units of ppmv atmospheric CO$_{2}$ emitted by
civilization per Joule of energy consumption. 

The modeling approach here is aimed at the simplest
of possible approaches. In reality, the carbon cycle is much more complicated
than can be easily justified by a linear sink model
\citep{Cox2000,Canadell2007}. That said, even the current magnitude of the
CO$_{2}$ sink is not well constrained \citep{LeQuere2003}. Given current
uncertainties, assuming a linear sink that is in line with current
observations appears to provide long-range forecasts of [CO$_{2}$]
that are in good agreement with far more sophisticated models. More detailed
discussion is presented in Sect.~\ref{sec:SRES} and Appendix~C.

From Eqs.~(\ref{eq:eta}) and~(\ref{eq:resilience}), the rate of return $\eta$
changes at a rate given by
\begin{equation}
\frac{d\eta}{dt}~=~\frac{d\beta}{dt}~-~\frac{1}{\rho}~\frac{d~\ln~\left[{\rm CO}_{2}\right]}{dt}\label{eq:detadt}
\end{equation}

Model trajectories in wealth~$C$ and atmospheric carbon dioxide concentration
evolve subject to initial conditions in {[}CO$_{2}${]}, $C$, $\beta$
and $\gamma$. Note that global production~$P$ is a diagnostic quantity
given by Eq.~(\ref{eq:PetaC}).

The prognostic CThERM model expressed by Eqs.~(\ref{eq:dlnCdt})
to~(\ref{eq:detadt}) is incomplete because it lacks prognostic equations for the
carbonization of the world's wealth $c$\,=\,$E/\left({\lambda}C\right)$
(Eq.~\ref{eq:ElamcC}) and the coefficient of nominal production $\beta$\,=\,$\hat{P}/C$
(Eq.~\ref{eq:beta}). A more sophisticated model will need to address the
evolution of these terms.\footnote{In principle, the evolution of $\beta$ is
governed by two factors, as illustrated in Fig.~\ref{fig:thermschematic}. As
civilization or any other system grows, it depletes known available energy
reservoirs; at the same time, it expands into new reservoirs that were
previously unavailable or unknown. Past bursts in growth in
$\eta$\,=\,$\beta-\gamma$ are seen to have occurred around 1880 and 1950, perhaps
due to a sudden increase in availability of important new oil reservoirs
\citep{GarrettCO2_2009}. Presumably the future availability of energy
reservoirs will influence the value of~$c$ as well \citep{Sorrell2010}.}

\begin{figure}
\vspace*{2mm}
\center\includegraphics[width=10cm]{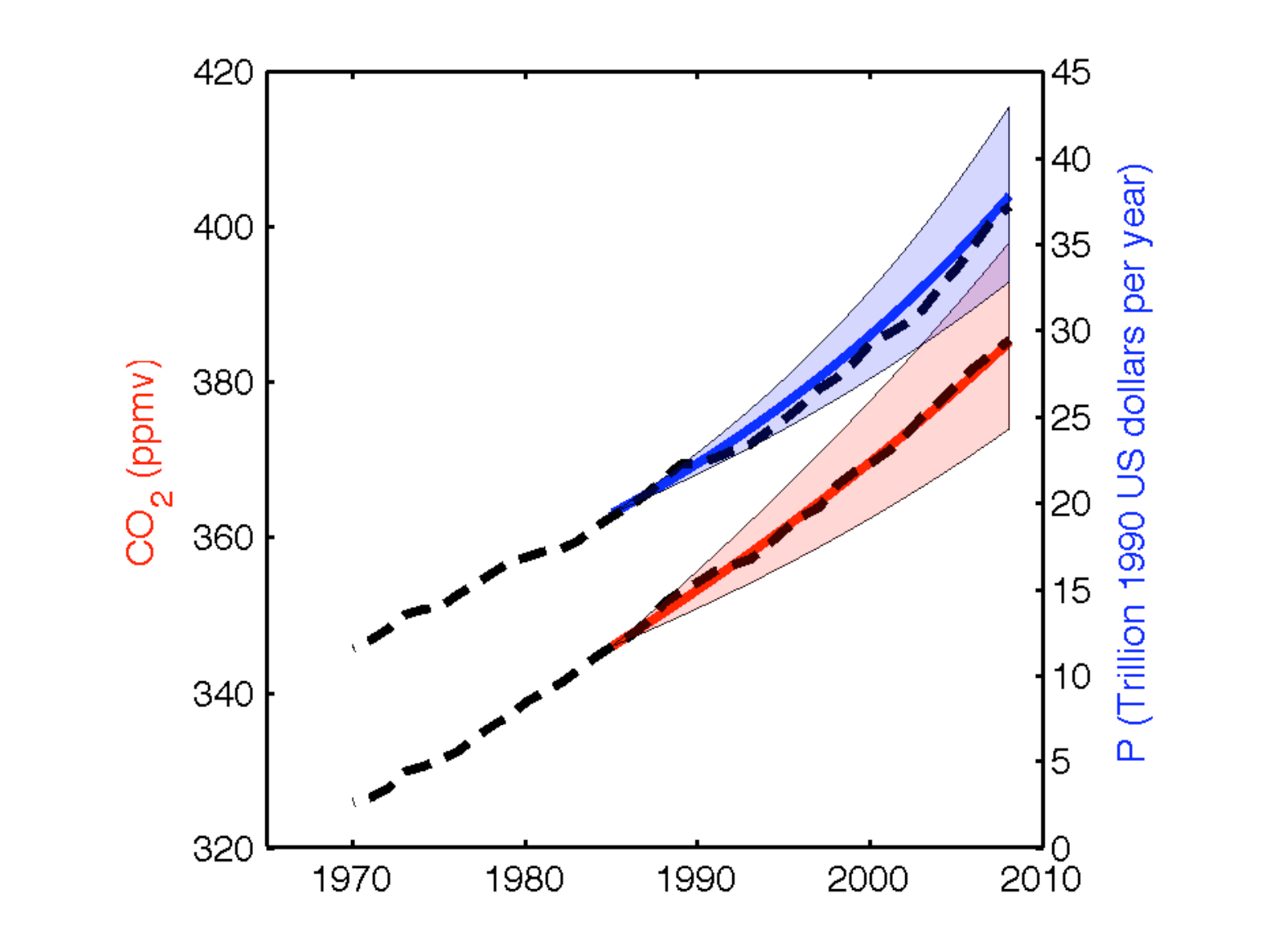}
\caption{Based on the CThERM model given by Eqs.~(\ref{eq:dlnCdt}) to~(\ref{eq:detadt}), hindcast trajectories and associated
uncertainty estimates for the period 1985 to 2008 in a space of atmospheric
CO$_{2}$ concentrations (red) and global economic production (blue). Observed
statistics for the period 1970 to 2008 are shown by black dashed lines. The
model is initialized with observed conditions in 1985, and a linear trend in
the nominal production coefficient $\beta$ between 1970 and 1984.\label{fig:hindcast}}
\end{figure}

A hindcast simulation that illustrates the accuracy of the model framework
is shown in Fig.~\ref{fig:hindcast}. The hindcast is initialized
in 1985 and, based on results shown in Fig.~\ref{fig:etabetagamma},
it is assumed that $d\gamma$/\textit{dt}\,=\,0 and that $d\beta$/\textit{dt} evolves on
a linear trajectory that is consistent with what is observed for the
period between 1970 and 1984. A linear fit for $d\beta$/\textit{dt} during this initialization time period is 0.017\%\,yr$^{-1}$ per year with a 95\% confidence limit of
$\pm$0.01\%\,yr$^{-1}$ per year. A second source of uncertainty is associated
with the CO$_{2}$ sink coefficient $\sigma$, which is estimated
to have a value of 1.55\,$\pm$\,0.75\,\%\,yr$^{-1}$ (Appendix~B).

Figure~\ref{fig:hindcast} shows that, with these assumptions, the mid-range of
hindcasts over a 23~year period between 1985 and 2008 faithfully reproduces
both the timing and magnitude of observed changes in atmospheric CO$_{2}$
concentrations and global economic production~$P$. The implication is that,
even though the model that is used is extremely simple, it is nonetheless
able to make accurate multi-decadal forecasts for the coupled growth of the global
economy and atmospheric composition. Furthermore, it suggests some ability of
the model to explore thermodynamically constrained forecasts in a space of
$P$ and {[}CO$_{2}${]} for a range of hypothetical values of civiilization
resilience $\rho$ and decarbonization rates $-d~\ln~c$/\textit{dt}.

As discussed previously, there is no good guidance yet for what a suitable
choice for the resilience $\rho$ might be, and no prognostic model is
included here for forecasting the evolution of either carbonization~$c$ or
the nominal productivity coefficient $\beta$. Thus, while the CThERM model is
thermodynamically constrained, it can still only provide forecasts for a
range of hypothetical scenarios in these parameters. In what follows, two
main categories of scenarios are considered.

\subsection{Forecast scenario~A: no decarbonization}

The first scenario that is considered here is simply to assume that for the
remainder of this century, there will be no further decarbonization, and that the
coefficient of nominal production will remain stagnant ( i.e., \textit{d{c}/d{t}}\,=\,0 and
$d\beta$/\textit{dt}\,=\,0 ). Figure~\ref{fig:trajectories} shows examples of forecasts for
these conditions for the years between 2009 and 2100. Also shown for
historical reference are past measurements between 1\,AD and 2008 (Appendix~C).

For this scenario, a range of resilience sub-scenarios can be considered. If
civilization is so resilient that it is unaffected by elevated CO$_{2}$
levels, then the world economy $P$ sustains recent growth rates of 2.2\% per
year. By 2100, it increases by nearly an order of magnitude to a value of
nearly 300~trillion 1990\,dollars per year. The accumulated production of
wealth $C$\,$\equiv$\,$\int_{0}^{2100}Pdt'$ corresponds to an increase in rates of
energy consumption $a$\,=\,${\lambda}C$ from 16\,TW in 2008 to 126\,TW in year 2100.
Absent any decarbonization, the accumulated and accelerating emissions push
CO$_{2}$ levels above 1100\,ppmv.

If, however, civilization has an extremely low resilience to elevated
CO$_{2}$ levels, then the decay coefficient $\gamma$ increases by 5\%\,yr$^{-1}$
per CO$_{2}$ doubling. Eventually, the decay coefficient exceeds
the coefficient of nominal production $\beta$. In this case, economic
production shrinks due to the impacts of climate change. Well before the year
2100, the inflationary pressure exceeds 100\%: real GDP is negative and
civilization is in a phase of collapse. However, even in this scenario,
energy consumption rates peak at 89\,TW in 2056 and although they fall to 21\,TW
in year 2100, they still exceed current levels. Because rates of energy
consumption remain high, even with rapid and immediate civilization collapse,
CO$_{2}$ levels still continue their rise to approximately 600\,ppmv by year
2100.

What is perhaps most striking when looking at these forecasts is that we can
expect some extraordinarily rapid near-term changes in the global economy
and atmospheric composition. For any plausible resilience condition,
atmospheric CO$_{2}$ concentrations and civilization GWP will change by as much
in the next 40~years as they have in the past two thousand.

\subsection{Forecast scenario~B: rapid decarbonization}

Although there is no evidence that civilization is in fact decarbonizing
\citep{Raupach2007}, one can imagine for the sake of illustration a second
forecast scenario shown in Fig.~\ref{fig:trajectories_decarbon} in which
$\beta$ stays constant, but the carbonization of civilization $c$ drops
extremely rapidly. Supposing that carbonization $c$ halves in just 50 years,
the value of~$c$ ends up 73\% lower in 2100 than it is at present. This is
highly imaginary, of course. If nothing else, no consideration is made here
of the costs of decarbonizing that would be involved. These would presumably
act to lower $\beta$ and be an inflationary pressure themselves (Eq.~\ref{eq:inflation}).
However, it is worth considering because, for one, it
illustrates how extremely rapid decarbonization would need to be to lower
CO$_{2}$ concentrations to something that only moderately exceeds the
450\,ppmv levels that might be considered to be ``dangerous''
\citep{HansenDangerous2007}. If civilization's resilience to climate change
is extremely low, then only a combination of rapid civilization collapse and
high decarbonization comes close to achieving a 450\,ppmv goal. Otherwise, if
civilization's resilience to climate change is extremely high, then emissions
increase from 3.95\,ppmv equivalent per year in 2008 to 8.64\,ppmv per year in
2100.

The reason why even rapid decarbonization still corresponds with increasing
emissions rates is that it has the side benefit of aiding economic
health and growth. By slowing growth in CO$_{2}$ concentrations,
the worst impacts of future climate change are constrained. Energy
consumption is fundamentally linked to the size of civilization through
the constant $\lambda$ (Eq.~\ref{eq:lambda}). Thus, any improvement
to economic wealth corresponds to increased energy consumption and
more rapid growth in CO$_{2}$ emissions (Eq.~\ref{eq:dlnEdt}).

It is counter-intuitive, but comparing two scenarios with very low resilience
to climate change, energy consumption rates rise about twice as fast with
rapid decarbonization as with no decarbonization. The reason is that
decarbonization aids society health by limiting global warming. Better health means greater energy
consumption, which then leads to a partial offset of any environmental gains that came
from decarbonizing in the first place.

\vspace*{2mm}
\subsection{Comparison with SRES scenarios\label{sec:SRES}}
\vspace*{2mm}

\begin{figure}
\vspace*{2mm}
\center\includegraphics[width=12cm]{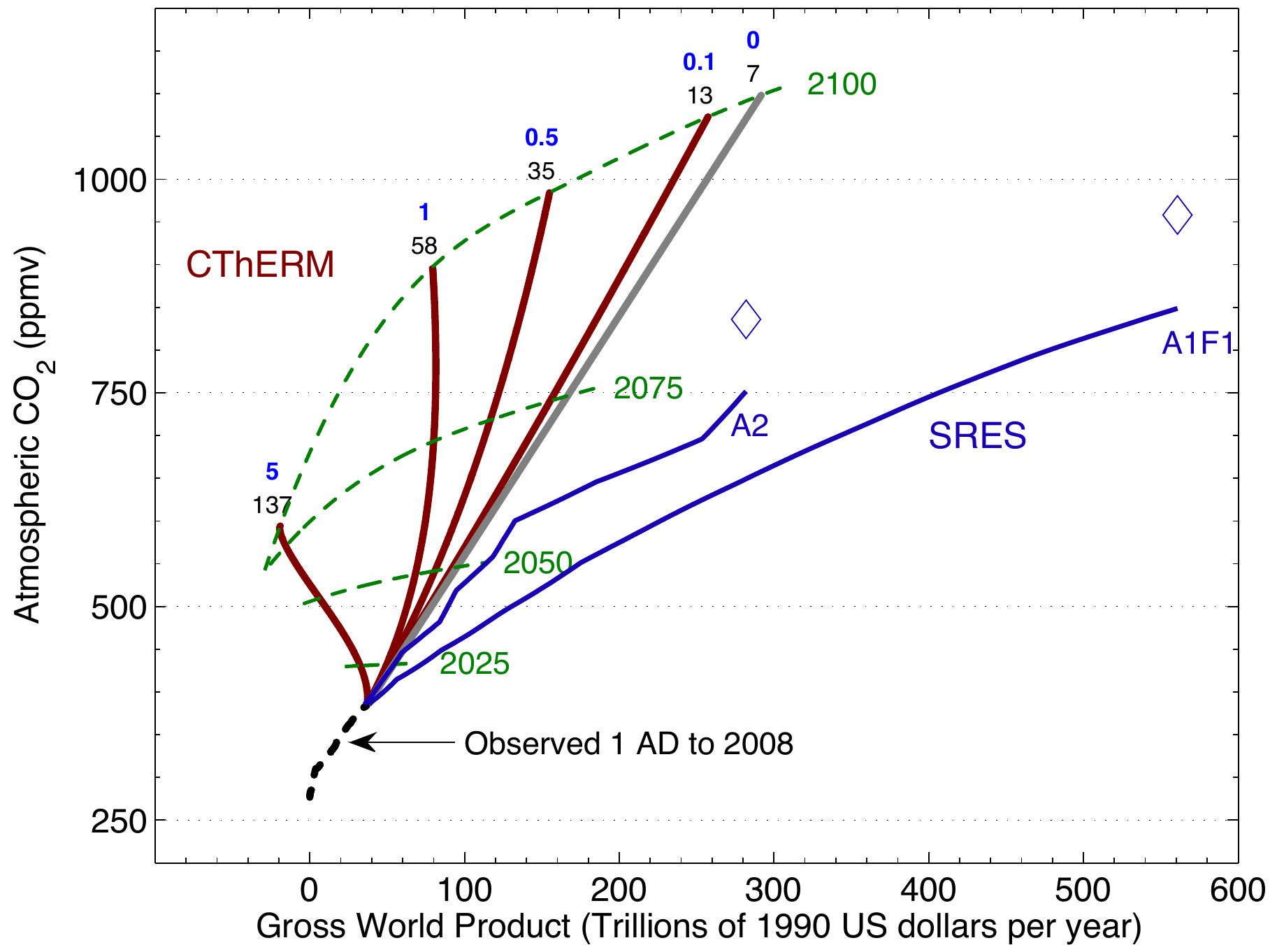}
\caption{As for Fig.~\ref{fig:hindcast}, except for
CThERM trajectories calculated out to 2100, with the model initialized with
conditions in 2008 and assuming that $d\beta$/\textit{dt}\,=\,0 and \textit{dc/dt}\,=\,0 for a range
of values of inverse resilience $1/\rho$ (blue numbers expressed in \%\,yr$^{-1}$
change in the decay coefficient $\gamma$ per CO$_{2}$ doubling).
Small numbers in black correspond to the calculated inflationary pressure
$i$\,=\,$\gamma/\beta$ (Eq.~\ref{eq:inflation}) in year 2100. Green dashed
lines represent the modeled year. Shown for comparison
are the IPCC SRES~A1F1 and A2~scenarios based on the CThERM linear sink model
for CO$_{2}$. CO$_{2}$ concentrations for these scenarios using the Bern carbon
cycle model are shown by blue diamonds. Historical data from 1\,AD to 2008 is
added for reference (see Appendix~C).\label{fig:trajectories}}
\end{figure}

\begin{figure}
\vspace*{2mm}
\center\includegraphics[width=12cm]{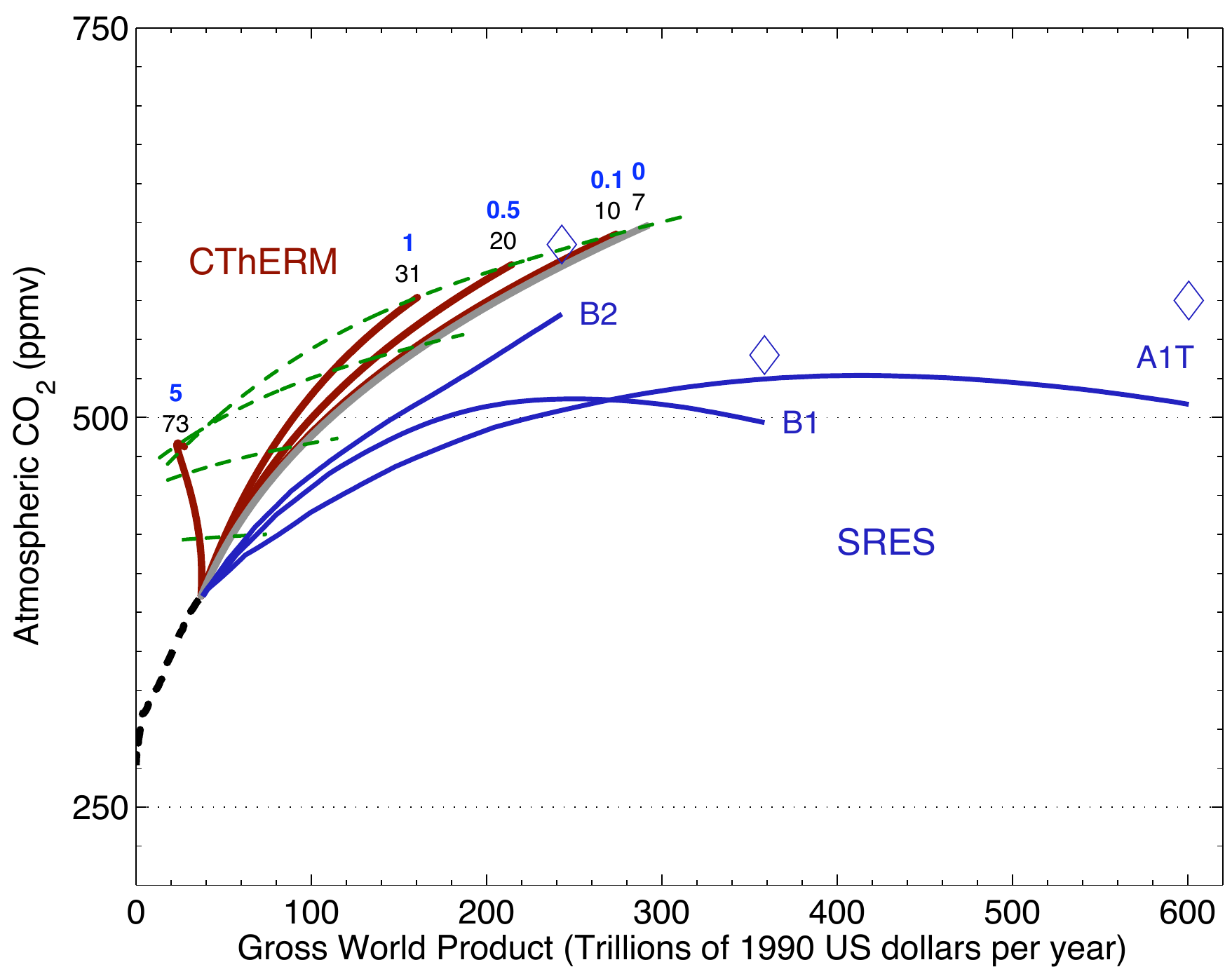}
\caption{As for Fig.~\ref{fig:trajectories} except that it is assumed that the value of
carbonization~$c$ has an assumed halving time of 50~years. For comparison,
the IPCC SRES trajectories that are considered are the A1T, B1 and B2
scenarios.\label{fig:trajectories_decarbon}}
\end{figure}

Figures~\ref{fig:trajectories} and~\ref{fig:trajectories_decarbon} include for
comparison's sake the phase space of $P$ and CO$_{2}$ concentrations that
are employed in several well-known IPCC Special Report on Emissions Scenarios
(SRES) illustrative marker scenarios \citep{IPCC_WG12007, Manning2010}. These scenarios
provide statistics through 2100 for global GWP in 1990\,MER~US dollars along
with global CO$_{2}$ emission rates from fossil fuel combustion. For the sake
of consistency with CThERM calculations, atmospheric CO$_{2}$ concentrations
are calculated from the second CThERM equation given by Eq.~(\ref{eq:ddCO2dt}).
Across the scenarios, calculated trajectories in CO$_{2}$ concentration
perturbations are lower than those presented in the IPCC Third Report for the
same emission rates, but calculated using the sophisticated ``Bern'' carbon
cycle model \citep{Joos1996}. Part of this discrepancy may be because no
consideration is made for the small additional perturbations in anthropogenic
CO$_{2}$ emissions that come from future non-fossil fuel sources. But also, no
account is made for possible future saturation of CO$_{2}$ sinks
\citep{LeQuere2007}. Regardless, the agreement is still sufficiently
favorable to support using the extremely simple CO$_{2}$ sink model in
Eq.~(\ref{eq:ddCO2dt}) as an accessible, if conservative, substitute for the more
sophisticated approaches used by the IPCC.

The comparisons between the CThERM and SRES scenarios are grouped according
to whether or not decarbonization is included in the forecasts. CThERM
trajectories in Fig.~\ref{fig:trajectories} include no decarbonization, and
are paired with the A1F1~and A2~scenarios. These two SRES storylines are both
based on a fossil-fuel reliant economy, while A1F1 has faster economic
growth. For contrast, the CTheRM trajectories in Fig.~\ref{fig:trajectories_decarbon}
do include decarbonization, and are paired
with the A1T, B1~and B2~scenarios. These storylines all include a switch to
less carbon intensive fuels, but with a range of speeds of economic
development.

Regardless of the precise scenario that is considered, there is a basic
difference between the CThERM forecasts and the SRES scenarios. Each SRES
scenario greatly underestimates how much atmospheric CO$_{2}$ concentrations
will rise for a given level of global GWP. Or, put another way, SRES
scenarios produce an unphysical overestimate of the wealth society can have
while maintaining CO$_{2}$ levels below some nominal threshold. For example,
the ``environmentally sustainable'' B1~scenario suggests that a CO$_{2}$
level below 500\,ppmv is plausible by the end of this century, while
maintaining a GWP of 360\,Trillion 1990\,US dollars per year. The CThERM
results suggest that this combination simply cannot happen because, even with
rapid decarbonization, sustaining this level of economic activity would
require too much energy consumption. It is only with rapid decarbonization
and civilization collapse that such CO$_{2}$ concentrations can be attained.

Perhaps the basic reason that there is a mismatch between the CThERM and SRES
scenarios is that the SRES scenarios are based on an assumption that
increases in energy efficiency will lower the amount of CO$_{2}$ emitted for a
given amount of economic activity. The thermodynamic and observational
analysis described here and in \citet{GarrettCO2_2009} indicate that the opposite should be expected to hold. From Eq.~(\ref{eq:etaefficiency}),
gains in efficiency $\epsilon$ accelerate CO$_{2}$ emissions by accelerating
civilization's capacity to access primary energy reservoirs. While, increasing efficiency may also lead to a higher GWP (Eq. \ref{eq:productionefficiency}), feedbacks in the economic system make it impossible to decouple the energy consumption from economic well-being.

\conclusions

This study builds on a key result presented in a prior article
\citep{GarrettCO2_2009}, that civilization wealth and global rates of primary
energy consumption are tied through a constant value of $\lambda$\,=\,9.7\,$\pm$\,0.3\,mW
per 1990\,US dollar. On this basis, a very simple prognostic
model (CThERM) is introduced for forecasting the coupled evolution of the
economy and atmospheric CO$_{2}$ concentrations. While the model in its basic
form has just three prognostic equations, it nonetheless provides accurate
multi-decadal hindcasts for global world production and atmospheric
concentrations of CO$_{2}$.

The much more sophisticated formulations commonly used in Integrated
Assessment Models can have hundreds of equations. In part this is required to
forecast regional variations of specific societal indicators such as
population or standard of living. The argument made here and in
\citet{GarrettCO2_2009} is that, at the global scales relevant to atmospheric
composition, such complexity is largely unnecessary. Both the global economy
and atmospheric CO$_{2}$ can be considered to be ``well-mixed'', and they
both are constrained by the global rate of primary energy
consumption.

One implication of this result is that global warming should
be expected to manifest itself economically as a growing gap between the nominal and inflation-adjusted GWP. Environmental pressures erode a material interface
that enables civilization to consume the primary energy resources it
requires. Normally, this erosion is more than offset by increasing access to
primary energy reservoirs; in fact, it is an increasing access to energy
supplies that has enabled a positive (and growing) inflation-adjusted gross
world product, and has led to the generally high standard of living we enjoy
today. However, in a global warming scenario, it can be expected that
environmental pressures will increase, and these will act to slow growth
in energy consumption. Fiscally, this will appear as an inflationary drag on
the growth of economic wealth. Ultimately, it is conceivable that it  will push civilization towards
an accelerating decline.

Another implication is that the commonly used IPCC SRES
scenarios make unphysical underestimates of how much energy will be needed to be consumed, and CO$_{2}$ emitted, to sustain prosperity growth. At the globally relevant scales, energy efficiency gains accelerate rather than reduce energy consumption gains. They do this by promoting civilization health and its economic capacity to expand into the energy reserves that sustain it. 

Reductions in CO$_2$ emissions can be achieved decarbonizing civilization's sources of fuel. But this has an important caveat. Decarbonization does not slow CO$_{2}$ accumulation by as much as might be anticipated because it also alleviates the potential rise in atmospheric CO$_{2}$ concentrations. If decarbonization leads to fewer climate extremes, then 
economic wealth is supported, and because wealth is tied to energy consumption through a constant, consumptive growth partly offsets the anticipated CO$_{2}$ emission reductions. Ultimately, civilization appears
to be in a double-bind with no obvious way out. Only a combination of
extremely rapid decarbonization and civilization collapse will enable
CO$_{2}$ concentrations to be stabilized below the 450\,ppmv
level that might be considered as ``dangerous'' \citep{HansenDangerous2007}.

\appendix
\section{\\ \\ \hspace*{-8mm} Thermodynamic accounting of decay}

\begin{figure}
\vspace*{2mm}
\center\includegraphics[width=12cm]{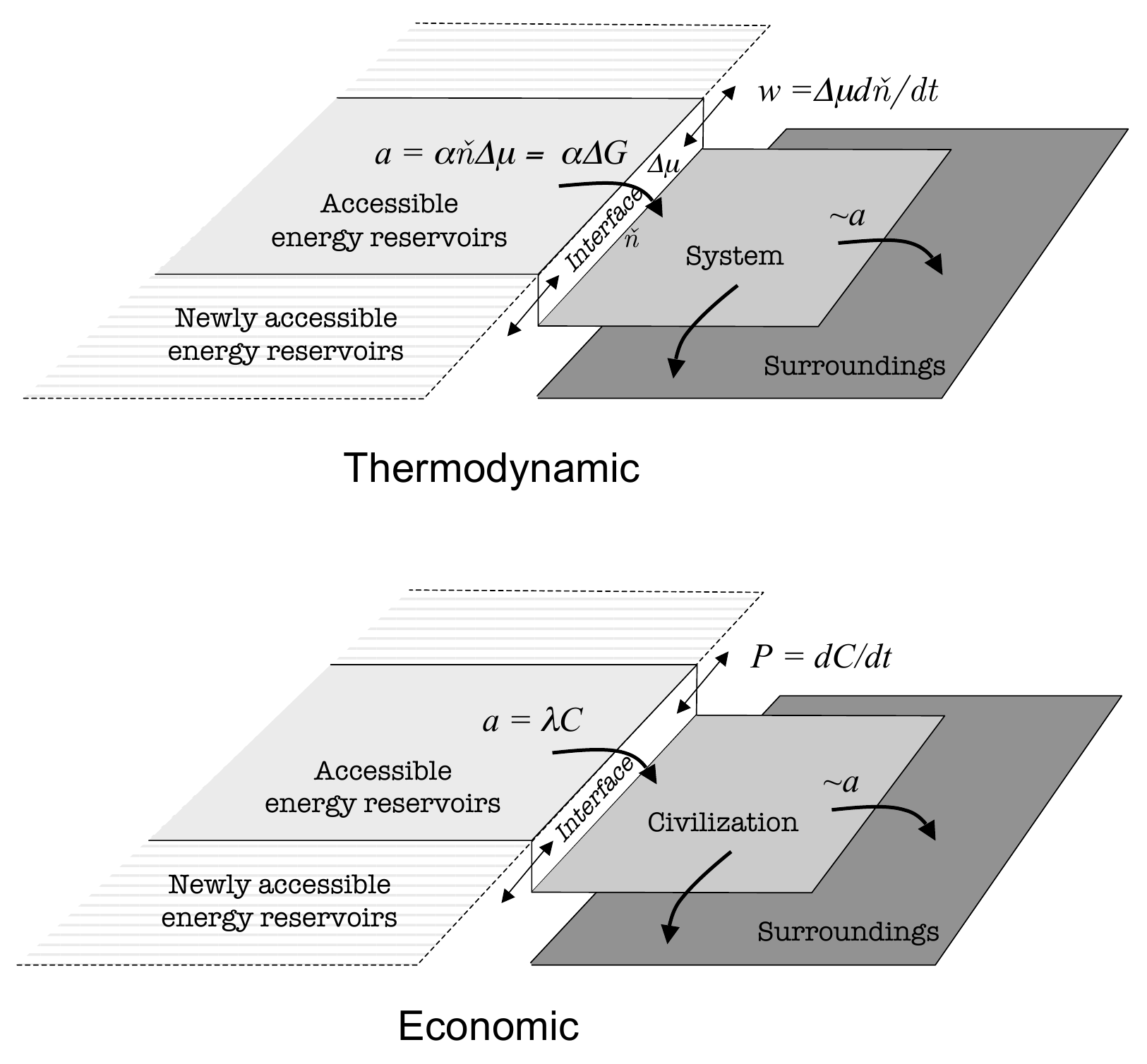}
\caption{Schematic for the thermodynamic evolution of an open system, and its hypothesized economic representation. Energy reservoirs, civilization, and its surroundings lie
along distinct constant potential surfaces. The number of material units $\breve{n}$ defining an interface between civilization and energy reservoirs determines the speed of downhill energetic flow at rate $a$, in proportion to a fixed specific potential difference $\Delta\mu$ and rate coefficient $\alpha$. The interface itself grows or shrinks at rate $w$ due to a net flux convergence into
civilization. In a positive feedback, interface growth at rate $d\breve{n}/dt$ expands energetic flows $a$ by extending civilization's access to previously inaccessible energy reservoirs at rate $da/dt$. Fiscally, wealth $C$ is proportional to both the interface size $\breve{n}$ and the rate of primary energy consumption $a$.  The GWP $P$ represents the net expansion of wealth at rate $dC/dt$ due to interface growth. $C$ and $a$ are linked through a constant $\lambda$. See the text and 
\citet{GarrettCO2_2009} for details.\label{fig:thermschematic}}
\end{figure}

The fiscal arguments for inflation discussed in Section \ref{economicdecay} can be
represented within the context of the generalized thermodynamic framework
illustrated in Fig.~\ref{fig:thermschematic}. Global wealth can be related to
thermodynamic flows through the constant $\lambda$, as framed by Eq.~(\ref{eq:lambda})
and validated through observations (Table~\ref{tab:Measured-values}). From Eq.~(\ref{eq:ddeltaGdt}), thermodynamic work
$w$ can be defined as the net growth rate in an interface with potential difference $\Delta{G}$ and number of material units $\breve{n}$. The interface
drives downhill thermodynamic flows at rate $a$\,=\,$\alpha{\Delta}G$\,=\,$\breve{n}\Delta\mu$, where $\Delta\mu$ is a fixed potential jump per unit matter.

Thus, from Eq.~(\ref{eq:rateofreturn}), thermodynamic work is done by civilization to expand the interface at rate
\begin{equation}
w~=~\frac{d\Delta{G}}{dt}~=~\Delta\mu\frac{d\breve{n}}{dt}\label{eq:work}
\end{equation}
Equation~(\ref{eq:rateofreturn}) dictates that, since $\lambda$ is a constant,
the rate of return $\eta$ applies equally to thermodynamic flows~$a$, the size
of the potential difference at the interface that drives flows ${\Delta}G$, and wealth~$C$. It follows that
the thermodynamic analog for the economic growth equations given by
Eqs.~(\ref{eq:sinksource}) or~(\ref{eq:dCdt}) is
\begin{equation}
\frac{d{\Delta}G}{dt}~=~\eta\Delta{G}~=~\hat{w}~-~\gamma~{\Delta}G~=~\left(\beta~-~\gamma\right)~{\Delta}G\label{eq:work decay}
\end{equation}
What this expresses is the details of how the interface shown in
Figure~\ref{fig:thermschematic} grows. Civilization grows by doing
{}``nominal'' work to stretch the interface driving flows outward at rate
$\hat{w}$\,=\,$\beta{\Delta}G$. By extension of Eq.~(\ref{eq:Palphaw}), nominal work is
the thermodynamic expression of nominal economic production through
\begin{equation}
\hat{P}~=~\frac{\alpha~\hat{w}}{\lambda}~=~\frac{\alpha}{\lambda}~\beta~{\Delta}G\label{eq:nominalprodtowork}
\end{equation}

However, it is only the ``real'' portion of work $w$\,=\,$d{\Delta}G/dt$ that
contributes to the net or real rate of interface growth: for real growth to
occur, nominal work $\beta{\Delta}G$ must be sufficiently rapid to overcome
natural decay $\gamma{\Delta}G$. Thus, real production~$P$
is related the size of the interface ${\Delta}G$ through
\begin{equation}
P~=~\frac{\alpha~w}{\lambda}~=~\frac{\alpha}{\lambda}~\left(\beta~-~\gamma\right)~{\Delta}G\label{eq:realprodtowork}
\end{equation}
Expressed in this fashion, real
economic production is a balance between two opposing thermodynamic forces
shown in Figure~\ref{fig:thermschematic}. There is an interface that connects
civilization to available energy reservoirs. Flow across this interface
arises from a consumption of primary energy resources. By consuming energy,
civilization both sustains its current size and does nominal work to
``stretch'' outward the size of the interface at rate $\beta$. As the
interface grows, it makes previously innaccessible or unknown reservoirs of
high potential energy (such as oil, coal, uranium, etc.) newly available. It
is by consuming and doing work that consumption accelerates.

However, this stretching only drives ``nominal'' growth. ``Real'' growth
takes into account environmental pressures that erode the interface at rate
$\gamma$. Such ``predation'' of civilization by the environment is due to a
loss of matter as things fall apart. There are many forms of material loss.
Photons are radiated through thermal heat loss; mass falls down due to
gravitation, and electrons are redistributed due to chemical reactions. What
matters from civilization's perspective is that this constant loss of
material hinders gains from nominal work $\hat{w}$. This slows the growth of
the interface ${\Delta}G$\,=\,$\Delta\mu\breve{n}$ that drives flows, and consequently it dampens
growth in energy consumption~$a$ and wealth~$C$. Due to material loss, only
net or real work is done at rate~$w$.

Thus, from Eqs. \ref{eq:nominalprodtowork} and \ref{eq:realprodtowork}, the thermodynamic form of the expression for economic inflation given by Eq. \ref{eq:inflation} is 
\begin{equation}
i~=~\frac{\int_{t}^{t+{\Delta}t}~\left(\hat{P}~-~P\right)~dt'}{\int_{t}^{t+\Delta
t}\hat{P}~dt'}~=~\frac{\int_{t}^{t+{\Delta}t}~\left(\hat{w}~-~w\right)~dt'}{\int_{t}^{t+\Delta
t}\hat{w}~dt'}~=~\frac{\int_{t}^{t+{\Delta}t}~\gamma~\Delta{G}dt'}{\int_{t}^{t+\Delta{t}}~\beta~\Delta{G}dt'}
~=~\frac{\left\langle\gamma\right\rangle}{\left\langle\beta\right\rangle}\label{eq:workinflation}
\end{equation}

As a note, a direct link can be drawn here to the increasingly popular concept of Energy Returned on Energy Invested (EROI)  \citep{Murphy2010}. The dimensionless EROI factor expresses how much energy society is able to recoup for consumption, relative to the amount of energy it must expend to access the energy. A point that is sometimes made is that the EROI is declining as new oil reserves become increasingly difficult to discover.

Here, the real rate of doing work $w$ is defined as the expansion in the potential gradient $\Delta{G}$, where the potential drives the flows of energy to society at rate $a$. Real work expands energy consumption at rate $da/dt~=~\alpha{w}~=~\alpha{d\Delta{G}/dt}$ (Eq. \ref{eq:dadt}). From Eq. \ref{eq:work decay}, civilization expansion is positive if there is a convergence of flows and the amount of potential energy that must be {}``expended'' in an effort to grow civilization energy consumption $\int_{t}^{t+\Delta{t}}\alpha\gamma\Delta{G}~dt'$ is less than the increase in the amount of potential energy that becomes newly available to be {}``consumed'' $\int_{t}^{t+\Delta{t}}\alpha\beta\Delta{G}~dt'$. Thus,  the EROI concept is expressible thermodynamically as  
\begin{equation}
\textrm{ EROI }~=~\frac{{\textrm{ Energy Consumption Gain }}}{{\textrm{ Energy Expenditure }}}~=~\frac{\int_{t}^{t+\Delta{t}}\hat{w}~dt'}{\int_{t}^{t+\Delta{t}}(\hat{w}-w)~dt' }~=~\frac{\int_{t}^{t+\Delta{t}}\alpha\beta\Delta{G}~dt'}{\int_{t}^{t+\Delta{t}}\alpha\gamma\Delta{G}~dt' }~=~\frac{\left\langle\beta\right\rangle}{\left\langle\gamma\right\rangle}\label{EROI}
\end{equation}
Since this is just the inverse of Eq. \ref{eq:workinflation}, the EROI is inversely tied to inflationary pressures through 
\begin{equation} 
i~=~ \frac{1}{\textrm{ EROI }}\label{inflationEROI}
\end{equation}
For example, a global EROI of 20 calculated over a given time period $\Delta{t}$, corresponds to a corresponding inflationary pressure of 5\%. If global civilization ever gets to the point that it expends as much energy during the extraction process as it is able to consume in return, then the inflationary pressure is 100\%, the EROI value is unity and civilization wealth $C$ is on the verge of tipping into collapse.  Any expansion work that civilization does serves only to maintain a standstill.  

\appendix
\section{\\ \\ \hspace*{-8mm} Parameterization of a linear sink term for CO$_{\vec 2}$}

A portion of the anthropogenic CO$_{2}$ that is accumulating in the
atmosphere has a concurrent sink to the land and oceans, both from
natural processes and changes associated with land-use. The nature
of the sink is complex, and depends on multiple processes with timescales
that vary by orders of magnitude. Detailed assessments of the magnitude,
trends, and uncertainties in the airborne fraction of CO$_{2}$ emissions~$E$
are provided by \citet{Canadell2007}, and ideally would require
a fully coupled Earth System Model \citep{Gent2009}.

For the sake of simplicity of argument, the carbon dioxide sink is assumed
here to be a linear function of the disequilibrium in atmospheric CO$_{2}$
concentrations~$C$. To see why this might not be as terrible a choice as it
might initially appear, consider the simple analytic representation of a
detailed carbon cycle model \citep{Joos1996}, which shows that a small pulse
of CO$_{2}$ into the atmosphere decays over multiple timescales
\citep{HansenDangerous2007}:
\begin{equation}
{\rm CO}_{2}~\left(\%\right)~=~18~+~14~e^{-t/420}~+~18~e^{-t/70}~+~24~e^{-t/21}~+~26~e^{-t/3.4}\label{eq:CO2model}
\end{equation}
This formulation points to multiple sink coefficients with decay weighted
towards shorter timescales, meaning that recent, faster emissions decay at a
more rapid rate than older, slower contributions. Thus, super-exponential
(i.e.~the exponent of an exponent) emissions growth would progressively bias
the instantaneous, or effective, value of the sink rate to ever shorter
timescales. If, however, CO$_{2}$ emissions grow nearly at a logarithmically
constant rate, then the linear sink rate for these CO$_{2}$ emissions
$\sigma$ (Eq.~\ref{eq:ddCO2dt}) should be approximately constant with time.

Currently, CO$_{2}$ emissions growth is nearly exponential, so assuming that
$\sigma$ is nearly constant, its value can be estimated by combining data for
the ocean and land sink \citep{LeQuere2003} with an assumed pre-industrial
equilibirum concentration of 275\,ppmv \citep{Wigley1983}. This leads to an
approximate value for $\sigma$ of 1.55\,$\pm$\,0.75\% per annum, corresponding to
a sink timescale of about 65~years (Table~A1).

The above framework neglects changes in CO$_{2}$ sinks that might be expected
to change in the future if, for example, there is saturation of the ability
of the earth's ecosystems and oceans to uptake carbon
\citep{Cox2000,LeQuere2007}. Certainly the systems involved are complex and
this adds to the difficultly of making confident quantification of future
behavior. Simply estimating a constant linear sink coefficient for
atmospheric CO$_{2}$ based on recent observations is aimed more at simplicity
than accuracy, and certainly more sophisticated forecasts than presented here
could implement some functional dependence for
$\sigma\left(\left[\Delta{\rm CO}_{2}\right]\right)$. However, given that
there are such large uncertainties on even the current magnitude of the
CO$_{2}$ sink, assuming a linear sink coefficient seems adequate until
estimates of carbon fluxes can be further constrained.

\section{\\ \\ \hspace*{-8mm} Historical records of economic production and CO$_{\vec 2}$ concentrations}

Historical measurements of atmospheric CO$_{2}$ perturbations from an assumed
baseline of 275\,ppmv are shown in Fig.~A1. Measurements
come from a combination of in-situ measurements from Mauna Loa
\citep{KeelingWhorf2005}, and Antarctic ice core data from the EPICA Dome~C
\citep{Fluckiger2002} and the Law Dome \citep{Etheridge1996}. These data are
compared to a time series for measurements of global world production that is
derived from a combination of statistics in 1990 market exchange rate dollars
available since 1970 \citep{UNstats} and more intermittent, long-term
historical estimates for the years~0 to~1992 derived by \citep{Maddison2003}.
For details see \citet{GarrettCO2_2009}. Although it is unclear exactly why,
the two millennia data in production~$P$ and and {[}$\Delta$CO$_{2}${]} are
well-represented by a remarkably simple power-law fit that accounts for 90\%
of the variance ($r$\,=\,0.952)
\[
\left[\Delta{\rm CO}_{2}\right]~=~2.5~P^{0.61}
\]
The results suggest a fairly
long term anthropogenic influence on atmospheric composition. It might be
tempting to infer from these data that CO$_{2}$ measurements at Mauna Loa
could be used to gauge the size of the global economy. However, obviously the
observed relationship between {[}$\Delta$CO$_{2}${]} and $P$ must break down
sometime in the future. $P$ is an instantaneous quantity, whereas CO$_{2}$
perturbations decay over timescales of hundreds to thousands of years (Eq.~\ref{eq:CO2model}).

\begin{acknowledgements}
Writing of this article benefited from a helpful review by Carsten Hermann-Pillath, discussions about Economics with Steve~Bannister, Harry~Saunders, Clint~Schmidt, Todd Kiefer, and Richard~Fowles, and Earth System Models with Court~Strong. This research was funded in part by the
Ewing~Marion~Kauffman Foundation whose views it does not represent.
\end{acknowledgements}

\bibliographystyle{/Users/tgarrett/LyX/STYFILES/EGU/egu}
\bibliography{References}

\clearpage

\begin{table}
\caption{Measured values for the global rate of energy consumption~$a$ (TW),
global wealth~$C$ (trillion 1990\,US\,\$), CO$_{2}$ emissions rates $E$
(ppmv atmospheric equivalent\,yr$^{-1}$), the hypothesized constant parameter $\lambda$ (mW per
1990\,US\,\$) and ${\lambda}c$ (ppmv\,yr$^{-1}$ per 10$^{15}$\,1990\,US\,\$)
where $c$\,=\,$E/a$.\label{tab:Measured-values}}
\vskip4mm
\begin{center}
\begin{tabular}{lrrrrrrrrr}
\tophline
& 1970 & 1975 & 1980 & 1985 & 1990 & 1995 & 2000 & 2005 & 2008 \\
\middlehline
$a$ (TW) & 7.2 & 8.4 & 9.6 & 10.3 & 11.7 & 12.2 & 13.2 & 15.3 & 16.4 \\
$C$\,=\,$\int_{0}^{t}P\left(t'\right)dt'$ & 821 & 884 & 960 & 1048 & 1151 & 1266 & 1398 & 1536 & 1656 \\
$\lambda$\,=\,$a/C$ & 8.8 & 9.4 & 10.0 & 9.8 & 10.2 & 9.6 & 9.4 & 9.9 & 9.9 \\
$E$ & 1.91 & 2.17 & 2.50 & 2.56 & 2.88 & 2.99 & 3.16 & 3.74 & 4.00 \\
${\lambda}c$\,=\,$E/C$ & 2.3 & 2.4 & 2.6 & 2.4 & 2.5 & 2.4 & 2.3 & 2.4 & 2.4 \\
\bottomhline
\end{tabular}
\end{center}
\end{table}

\clearpage

\begin{table}
{\bf Table~A1.}~{Estimates of the annual ocean and land net sink
for carbon (in Pg\,C\,yr$^{-1}$), including those associated with land-use
changes \citep{LeQuere2003}, the total sink (in ppmv CO$_{2}$\,yr$^{-1}$), the
decadal mean value of the carbon dioxide disequilibrium above 275 ppmv
$\left[\Delta{\rm CO}_{2}\right]$, and the associated linear sink
coefficient $\sigma$ (\%\,yr$^{-1}$). For convenience, the total sink is
expressed in units of ppmv atmospheric CO$_{2}$ per year through division by
the total atmospheric mass \citep{Trenberth1981}, such that 1\,ppmv
CO$_{2}$\,=\,2.13\,Pg emitted carbon.\label{tab:sinkrates}}
\vskip4mm
\begin{center}
\begin{tabular}{lccrcc}
\tophline
& Ocean sink & Land sink & \multicolumn{1}{c}{Total sink} &  $\left[\Delta{\rm CO}_{2}\right]$ & $\sigma$ (\%\,yr$^{-1}$) \\
& & & (in ppmv\,CO$_{2}$\,yr$^{-1}$) & & \\
\middlehline
1980s & 1.8\,$\pm$\,0.8 & 0.3\,$\pm$\,0.9 & 1\,$\pm$\,0.6 & 70 & 1.4\,$\pm$\,0.9 \\
1990s & 1.9\,$\pm$\,0.7 & 1.2\,$\pm$\,0.8 & 1.5\,$\pm$\,0.5 & 85 & 1.7\,$\pm$\,0.6 \\
\bottomhline
\end{tabular}
\end{center}
\end{table}

\clearpage

\clearpage

\clearpage

\clearpage

\clearpage

\clearpage

\clearpage

\clearpage

\begin{figure}
\vspace*{2mm}
\begin{center}
\includegraphics[width=12cm]{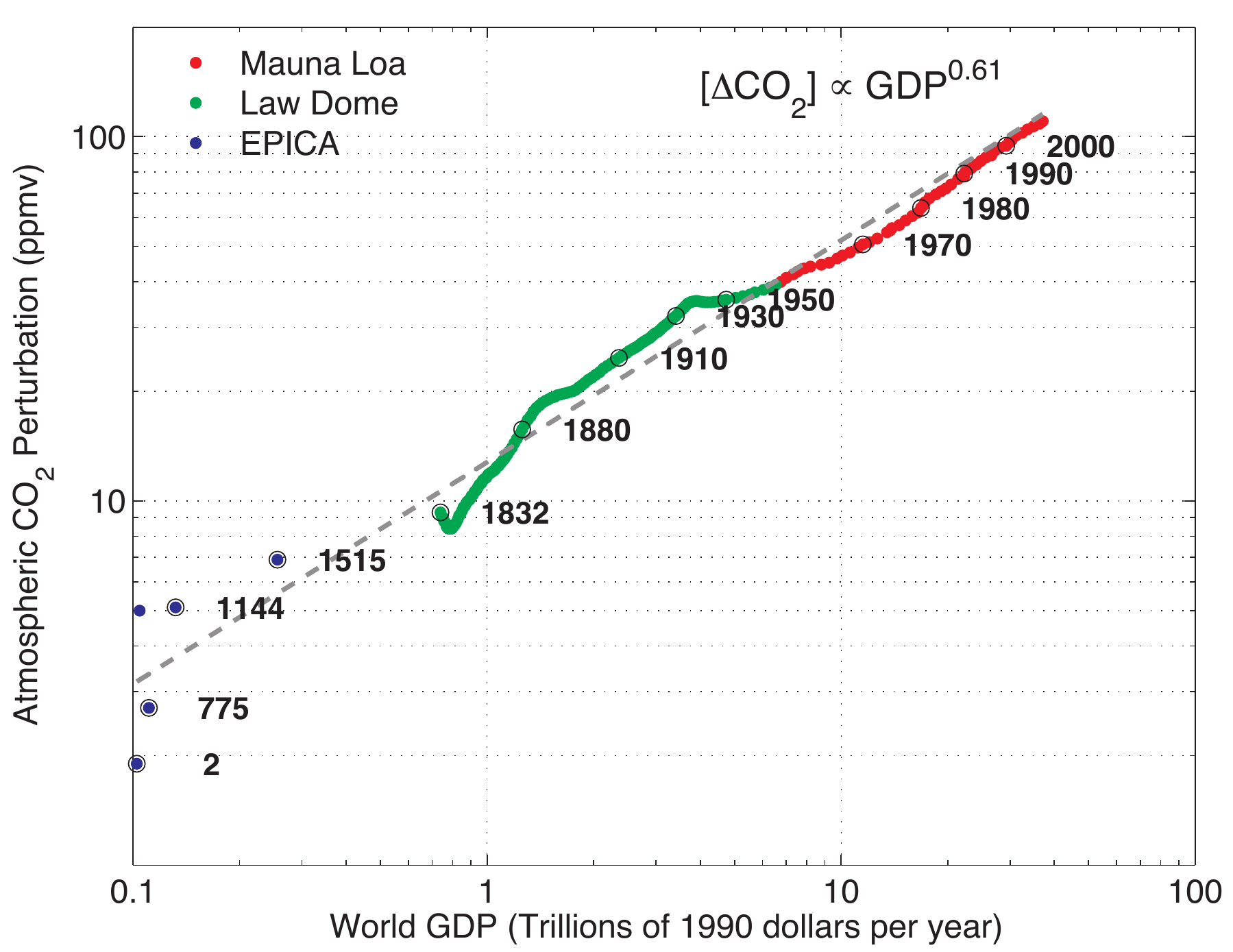}
\end{center}
{\bf Fig.~A1.}~{Measured perturbations in atmospheric CO$_{2}$
concentrations from a baseline of 275\,ppmv, compared with historical
estimates of global GDP in inflation adjusted 1990 dollars, with associated
year markers, and a linear fit to the data.\label{fig:GDPCO2past}}
\end{figure}

\end{document}